\documentclass[sigconf]{acmart}

\usepackage{listings}
\usepackage{dblfloatfix} 
\AtBeginDocument{%
  }

\setcopyright{acmlicensed}
\copyrightyear{2018}
\acmYear{2018}
\acmDOI{XXXXXXX.XXXXXXX}

\acmConference[Conference acronym 'XX]{Make sure to enter the correct
  conference title from your rights confirmation emai}{June 03--05,
  2018}{Woodstock, NY}
\acmISBN{978-1-4503-XXXX-X/18/06}

\begin{document}

\title{UIClip: A Data-driven Model for Assessing User Interface Design}

\author{Jason Wu, Yi-Hao Peng, Amanda Li, Amanda Swearngin, Jeffrey P. Bigham, Jeffrey Nichols}
\email{{jsonwu,yihao,xal}@cmu.edu, {aswearngin, jbigham, jwnichols}@apple.com}
\affiliation{%
  \institution{Carnegie Mellon University, Apple}
  % \streetaddress{Leave Anonymous}
  % \city{Leave Anonymous}
  % \state{Leave Anonymous}
  \country{USA}
}
% \author{Jason Wu}
% \email{jsonwu@cmu.edu}
% \affiliation{%
%   \institution{HCI Institute, Carnegie Mellon University}
%   \streetaddress{5000 Forbes Ave}
%   \city{Pittsburgh}
%   \state{PA}
%   \postcode{15213}
% }
% \author{Jeffrey Bigham}
% \email{jbigham@cs.cmu.edu}
% \affiliation{%
%   \institution{HCI Institute, Carnegie Mellon University}
%   \streetaddress{5000 Forbes Ave}
%   \city{Pittsburgh}
%   \state{PA}
%   \postcode{15213}
% }

% \author{Jeffrey Nichols}
% \email{jwnichols@apple.com}
% \affiliation{%
%   \institution{Apple Inc.}
%   \city{Seattle}
%   \state{WA}

\renewcommand{\shortauthors}{Wu et al.}


% aalto interface metrics, GPT-4, Sketchplore, Scout

\begin{abstract}
User interface (UI) design is a difficult yet important task for ensuring the usability, accessibility, and aesthetic qualities of applications.
In our paper, we develop a machine-learned model, UIClip, for assessing the design quality and visual relevance of a UI given its screenshot and natural language description.
To train UIClip, we used a combination of automated crawling, synthetic augmentation, and human ratings to construct a large-scale dataset of UIs, collated by description and ranked by design quality.
Through training on the dataset, UIClip implicitly learns properties of good and bad designs by \textit{i)} assigning a numerical score that represents a UI design's relevance and quality and \textit{ii)} providing design suggestions.
In an evaluation that compared the outputs of UIClip and other baselines to UIs rated by 12 human designers, we found that UIClip achieved the highest agreement with ground-truth rankings.
Finally, we present three example applications that demonstrate how UIClip can facilitate downstream applications that rely on instantaneous assessment of UI design quality: \text{i)} UI code generation, \textit{ii)} UI design tips generation, and \textit{iii)} quality-aware UI example search.

\end{abstract}



\keywords{UI Modeling; UI Design Assessment; Dataset}


\maketitle

\section{Introduction}
% not sure about the intro framing at the moment. seems to focus on speed but i think it would be better if it focused on the difficulty of assessing good designs in general.
% User interface (UI) design is a difficult task that requires significant experience and effort.
% \textcolor{red}{talk about how design is hard for computational approaches as a part of the movitation. and we need a way of encoding design knowledge}

% potentially useful text for alternative framing
% % Many concepts in HCI and design are qualitative or subjective.
% However, there are many benefits to quantifying these concepts for machine-assisted and automated de.
% \textcolor{red}{this sounds like it could go in intro probably}
% While many design concepts are described subjectively, there has been a long history of research that attempts to quantify them.
% In \textit{The Sciences of the Artificial}~\cite{simon2019sciences}, author Herb Simon offers an early formulation of design as an logical search process \emph{---} one that maximizes the ``utility'' of an object or artifact given parameters and constraints. Later work has successfully applied these principles to automatically design and generate UIs \cite{oulasvirta2022computational,gajos2004supple} also through optimization or search processes.
% A important component of this framework, especially for UIs, is the ``scoring function" which provides accurate and rapid estimation of design quality.

What makes a good user interface (UI)?
% not sure about this sentence
It is hard to comprehensively articulate what separates a good UI design from a bad one, and the task of UI design is challenging even for experts with years of training and practice.
Guidelines exist that list some general principles \cite{nielsen1992finding,shneiderman2016designing}, but they are often insufficient or difficult to operationalize, especially for novice designers.
Because of this, many application UIs today contain common design problems, which can negatively impact usability, accessibility, and design aesthetics.

% guidelines: CRAP, ...

The most holistic method of evaluating UIs is \textit{usability testing}, which can uncover UI design flaws, accessibility problems, and software bugs, but it is generally a time-consuming and costly process.
% Usability testing generally involves asking participants (\textit{e.g.,} potential users) to interact with a product or prototype to gauge it's quality.
% Despite the potential to provide substantial subjective feedback and deep insights into app design, the process of usability testing has several drawbacks.
% Usability testing is time-consuming, costly, and potentially non-reproducable, since it depends on the number and make-up of participants. % cite NNgroup's paper on number of participants needed and tradeoff
% lo-fi, paper prototypes, wireframing
% Sketchplore, heuristic evaluation
Approximate assessments, such as \textit{heuristic evaluation}, rely on experts applying a set of pre-defined principles to rapidly identify potential problems and estimate overall UI quality.
However, even these abbreviated strategies can be difficult to employ consistently or in the absence of a knowledgeable expert.
% computational methods are promising because

% Traditionally, these processes have relied on human experts because many concepts in HCI and design are qualitative or subjective.
To this end, computational methods have been developed to estimate the quality of UIs, taking into account factors such as visual aesthetics~\cite{miniukovich2015computation}, cognitive principles~\cite{oulasvirta2018aalto}, and context~\cite{oulasvirta2022computational}.
Because of their automated nature, they unlock new opportunities for UI design~\cite{swearngin2020scout,todi2016sketchplore} and evaluation~\cite{moran2018automated}.
However, most of these prior computational approaches are limited.
Some techniques apply objectives and metrics inspired by cognitive principles \cite{miniukovich2015computation,oulasvirta2018aalto}, such as visual complexity, layout quality, and color harmony to UI designs, but their outputs still require interpretation and cannot, for example, be used to compare the quality of two candidate designs.
Other approaches are toolkits that learn user-specific models for generating adaptive interfaces~\cite{gajos2004supple,gajos2005preference,gajos2007automatically}, and they also cannot be applied to more generalized UI design tasks.
% However, there are many benefits to quantifying these concepts for machine-assisted and automated de.
% \textcolor{red}{this sounds like it could go in intro probably}
% While many design concepts are described subjectively, there has been a long history of research that attempts to quantify them.
% In \textit{The Sciences of the Artificial}~\cite{simon2019sciences}, author Herb Simon offers an early formulation of design as an logical search process \emph{---} one that maximizes the ``utility'' of an object or artifact given parameters and constraints. Later work has successfully applied these principles to automatically design and generate UIs \cite{oulasvirta2022computational,gajos2004supple} also through optimization or search processes.
% A important component of this framework, especially for UIs, is the ``scoring function" which provides accurate and rapid estimation of design quality.

Our paper introduces a novel computational model, \textit{UIClip}, to assess the design quality of any UI from its screenshot.
% As its name suggests, 
UIClip is based on the well-known CLIP vision-language model~\cite{radford2021learning}, and it uses a natural language description of the UI coupled with a screenshot to assign a numerical score that estimates design quality.
CLIP, by default, is not well-suited for judging UI quality and relevance.
Therefore, to train UIClip, we developed a novel technique for synthetically generating a large-scale dataset of UIs ranked by design quality.
Our strategy takes existing UIs (e.g., web pages) and intentionally introduces design defects by modifying style and layout attributes. The process created pairs of original and ``jittered'' interfaces and allowed the models to learn the differentiation between these pairs. 
We used this method to generate 2.3 million pairs of UIs coupled with their quality-related descriptions. To align our model with real-world design preferences, we collected 1.2K ratings from professional designers on an extra UI set. These ratings were used to refine UIClip and validate the effectiveness of our model. 
% Our approach takes existing UIs (e.g., web pages) and intentionally introduces design defects by controllably modifying various style and layout properties, resulting in pairs of original and ``jittered'' interfaces, which a model is trained to distinguish. 
% We applied this method to create 2.3 million pairs of UIs and their related quality descriptions. Additionally, we gathered 1,200 UI ratings from designers to better align our model with real-world design preferences and to validate its effectiveness.
% summarize the way to incorporate designer ratings
% In a quantitative evaluation, we used UIClip and several other strong large vision-language models (LVLMs) baselines to rate a held-out set of UI screens and found that UIClip had the best performance on all tested tasks, beating baseline models several orders of magnitude larger.

We benchmarked UIClip with other large vision-language models (LVLM) by evaluating them on a held-out set of UI screens. We assess the models on three tasks, including design quality, improvement suggestions, and design relevance. The results showed that UIClip outperformed all other models in every task, despite being smaller in size.
% We benchmarked UIClip with various large vision-language models by evaluating a held-out set of UI screens. 
% We evaluated these models based on the key factors of quantitative UI assessment, including the design quality, suggestions for improvement, and relevance between different designs.
% The results showed that UIClip outperformed all the other models, even those several orders of magnitude larger in size, across every task we tested.
Finally, to demonstrate the utility of UIClip, we present three example applications that use our model to provide different types of computational UI design assistance: \textit{i)} quality-aware UI code generation, \textit{ii)} UI design suggestion generation, and \textit{iii)} quality-aware UI example retrieval.

To summarize, our work makes the following contributions:

\begin{enumerate}
    \item A large-scale dataset of UI designs and descriptions comprised of synthetic and human-generated design ratings.
    \item A computational model that scores UI screenshots based on relevance to a textual description and design quality.
    \item Three example applications that demonstrate how UIClip can be used to facilitate downstream applications: \textit{i)} a tool that improves the quality of UI code generated by LLMs, \textit{ii)} a tool that generates design recommendations for a UI screenshot, and \textit{iii)} a UI design search engine.
\end{enumerate}

To facilitate research in this area, we plan to release all the training code, data, and models.

% intro note:  maybe we would put somewhere saying that CLIP is relatively “small” multimodal model compared to all other alternatives (~88.3M + 38 M million parameters, depends on the variation we chose), so could have potential to apply to many applications, cuz technically our approach would also work to perform visual instruction-tuning for other (open) VLMs, but we chose one that could have broad applicability

% parameter refs for CLIP-related models: https://arxiv.org/pdf/2309.12314.pdf

\section{Related Work}
Our research builds upon existing work in UI design and evaluation by encoding UI design quality into computational models, enabling the models to serve as potential tools for UI design assessment. We review the literature in three relevant areas: UI design tools, UI evaluation, and machine learning-based quality metrics.

\subsection{UI Design Tools}
Embedding computational capabilities into UI design tools enables machines to computationally assess the design thus empowering designers to ideate, prototype, and iterate their work effectively.
% Early work such as SILK~\cite{landay1996silk} and DENIM~\cite{newman2003denim} empower designers to produce works through quick sketching. Damask introduced the pattern-based design creation flow, 
Early research like SILK~\cite{landay1996silk} and DENIM~\cite{newman2003denim} introduced quick sketching capabilities, making the design process more agile.
Damask \cite{lin2002damask} refined the creation process with its emphasis on pattern-based design, enhancing UI component reusability. 
The evolution continued with Smart Templates~\cite{nichols2004improving}, which provided designers with adaptable frameworks that intelligently adjusted to their needs, simplifying the design process.
Sikuli~\cite{yeh2009sikuli} built upon the thread of intelligent design tools by integrating image-based search functionalities, making it easier for designers to find and incorporate UI elements. 
As the field progressed, tools like Sketchplore~\cite{todi2016sketchplore} and Scout~\cite{swearngin2020scout} enabled designers to explore a wider array of design alternatives, encouraging creativity.  D.note~\cite{hartmann2010d} and Swire~\cite{huang2019swire} introduced interactive elements that incorporated user feedback directly into the design, enhancing user-centric approaches. 
The integration of deep learning into UI design tools marked a pivotal shift, starting with the use of datasets like RICO~\cite{deka2017rico} to inform model training. 
For instance, GUIComp~\cite{lee2020guicomp} is a tool that includes an autoencoder trained on the large-scale UI dataset to help find UI design examples for inspiration. In addition, the tool employed convolutional neural networks to evaluate the visual complexity of UI prototypes and pinpoint the main areas of interest.
Similarly, VINS~\cite{bunian2021vins} introduced a visual search framework powered by models trained on a more diverse annotated UI dataset, enabling designers to find similar visual UI designs across platforms. 
Our work builds upon existing work in computational UI design tools by building neural models to quantify UI design quality through language, and integrate the models into various UI design applications.

% REVISION TODO: 
% Incorporate more compute-ml-based metrics in tools (some works mentioned in the UI evaluation section)
% add HF websight, Stanford design2code (or maybe the next section?

% REVISION TODO: 

% Double-check that we include all related prior works (especially recent works)
% Remember to add Gajos's works
% gajos2005preference --- Preference elicitation for interface optimization
% Design2code (? From screenshot to code and preference assessment?

\subsection{UI Evaluation}

Traditional UI evaluation, initially rooted in heuristic evaluation and established guidelines~\cite{nielsen1994enhancing, jansen1998graphical}, has evolved significantly over time. 
The development of automated metrics marked a transition towards more objective and scalable UI assessments. 
Early work like ARNAULD~\cite{gajos2005preference} collected \textit{user preferences} about specific outcomes to automatically learn and tailor a cost function for UI assessment and adaptation. 
tLight~\cite{miniukovich2015computation} continued this vision and presented eight automatic metrics for evaluating graphical user interfaces' aesthetics, demonstrating their effectiveness on desktop and mobile platforms.
Progressing further, researchers also explored assessing the visual complexity of mobile user interfaces, establishing metrics that link visual complexity to perceived usability~\cite{riegler2018measuring}. This shift underscores a growing emphasis on quantifying user interface elements to predict usability outcomes.  
Moreover, integrating cognitive principles into UI evaluation is gaining traction, with metrics now considering harmony and attractiveness, aligning with how users perceive and organize visual information. For instance, the Aalto Interface Metrics (AIM) service~\cite{oulasvirta2018aalto} demonstrates how blending user perception and attention models can improve GUI design evaluation.
In recent developments, deep learning has been employed to model user interaction aspects like tappability~\cite{swearngin2019modeling, schoop2022predicting} and draggability~\cite{wu2023never}, marking a shift towards using neural modeling to enhance our understanding of user behaviors. Furthermore, with the rise of generative models, recent research applies Large Language Models (LLMs) to provide UI design feedback~\cite{duan2024generating}, illustrating how combining design knowledge parameterized in large pre-trained models with user input can be helpful for designers to improve the visual UI design. Our research builds on these advancements by linking UI design with quality-focused natural language descriptions, leveraging language as a tool for retrieval and feedback in design.

% Traditional UI evaluation mostly focused on heuristic evaluation and constructed guidelines.
% Several metrics have been developed for automatically evaluating UIs
% Many of these metrics focus on cognitive principles like gestalt, harmony. 
% % expand on this

% The Aalto Interface Metrics [x] is an example of a toolkit that combines several empirically-validated metrics together.

% With the advancement of generative models, recent work generates UI Design Feedback with LLMs~\cite{duan2023towards}

% % Full paper version --- Generating Automatic Feedback on UI Mockups with Large Language Models Peitong Duan, Jeremy Warner, Yang Li, Björn Hartmann CHI 2024 (To Appear)

% \vspace{-12px}

% REVISION TODO: 

% include more IQA type of work
% language modeling, visual quality estimation.
% Image Quality Assessment: IQA

% Grammatical Error Correction: A Survey of the State of the Art
% diffusion models, implicit learning

\subsection{Machine Learning-based Quality Metrics}

Learning scoring functions has been an important topic in many areas of machine learning. 
In the context of text-generation or machine translation, a popular class of text quality metrics involve the use of ``ground truth'' responses known as ``references.''
BLEU~\cite{papineni2002bleu} and later variants like ROGUE~\cite{lin2004rouge} and Meteor~\cite{banerjee2005meteor} were developed for other applications, such as summarization~\cite{giannakopoulos2011autosummeng}. % comet
% However, human-authored references may not be available for some domains. 
% ``Reference-free" metrics were developed that obviate the need for human annotations at test-time.
% Perplexity is a classic metric that approximates how likely a piece of text (often machine-generated) is to be sampled from a (often human-generated) corpus. % double check this
However, not all domains have access to human-authored references, leading to the development of "reference-free" metrics. Perplexity~\cite{jelinek1977perplexity}, for instance, is a classic metric used to estimate how likely a piece of text, often generated by a machine, is to come from a human-generated corpus.
More recently, direct human evaluations have been utilized to assess model-generated text~\cite{zheng2024judging}. This type of evaluation system often ask individuals to compare outputs from the same input text to determine which model-generated version they prefer.
In the realm of computer vision, numerous metrics have been devised to evaluate the quality of images produced by models, including the inception score~\cite{salimans2016improved}, the Fréchet Inception Distance (FID)~\cite{heusel2017gans}, and more recently the HyPE scores~\cite{zhou2019hype}.
Finally, there has been a class of evaluation methods aimed at multi-modal applications that concern both text and images.
CLIPScore~\cite{hessel2021clipscore} is a technique for assessing the quality of image captions by using the pre-trained OpenAI CLIP model. 
CLIP-IQA~\cite{wang2023exploring} further adopts CLIP to contrastively learn a function that evaluates images based on various quality attributes (e.g., brightness, colorfulness) and perceptual aspects (happy, scary).
TIFA~\cite{hu2023tifa} introduces a method where it asks and answers its own visual questions using large vision-language models. It then quantifies how well the text prompts and the images generated from those prompts align.
In our paper, we show that off-the-shelf vision-language models like CLIP often fall short in accurately analyzing UI screenshots, particularly when assessing UI design quality through language. To tackle this problem, we introduce a comprehensive UI design quality dataset that integrates both machine and human feedback. The collected data enables researchers to build and iterate their computational models using this quality-encoded dataset.

\begin{figure*}
    \centering
    \includegraphics[width=0.95\linewidth]{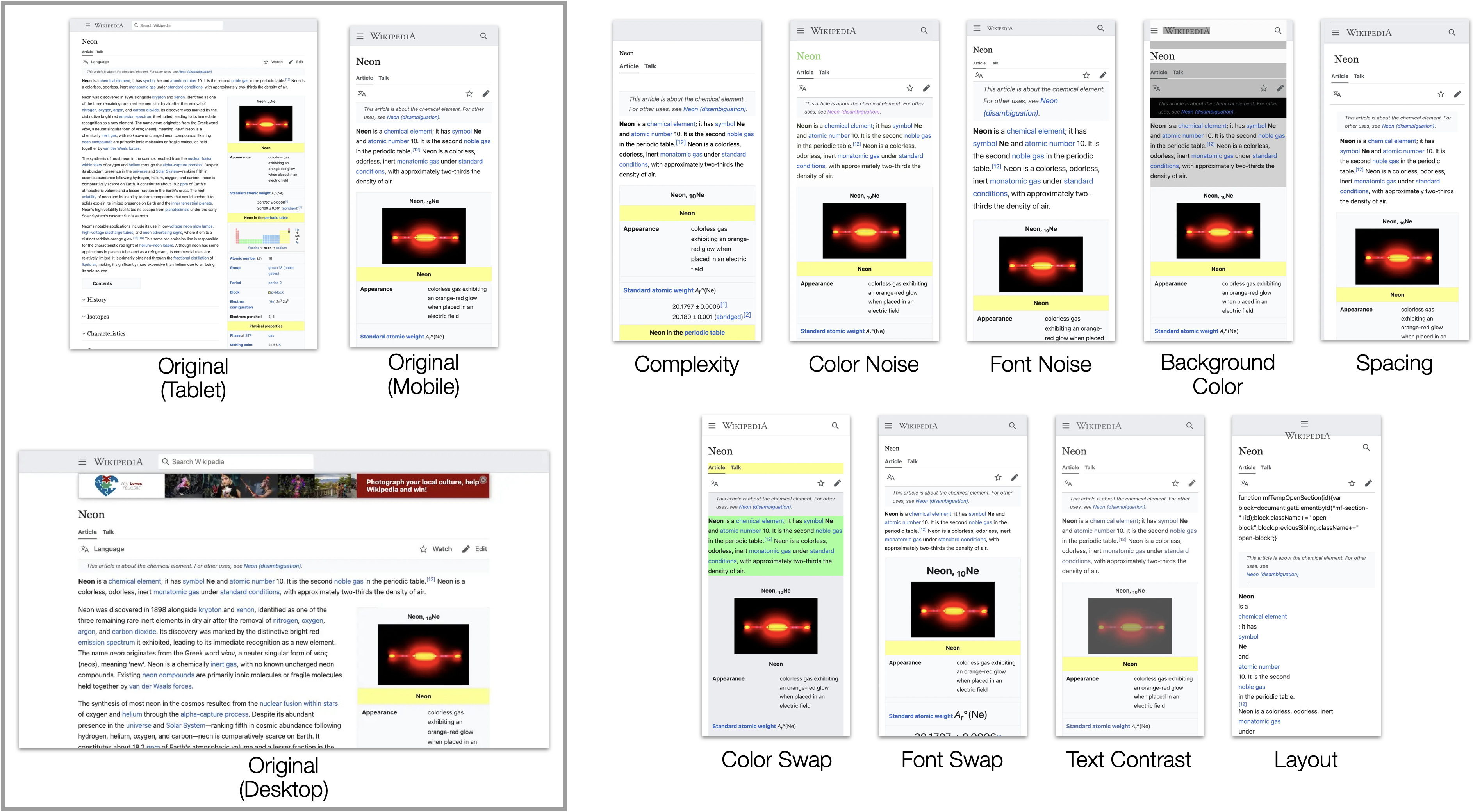}
    \caption{Our synthetic dataset was comprised of UIs that were processed by \textit{jitter functions} to introduce design defects. In this figure, we visualize the effect of each jitter function independently, although up to three jitter functions can be applied simultaneously. Our crawler captures screenshots for multiple devices (desktop, tablet, and mobile), but due to space constraints, we only show rendered mobile examples.}
    \label{fig:jitters}
    \Description[]{}
\end{figure*}

\section{Datasets for UI Design Quality}
While several UI datasets exist, they are annotated for other applications, such as element detection~\cite{deka2017rico,bunian2021vins}, natural language description~\cite{wang2021screen2words}, and app categorization~\cite{leiva2020enrico}.
Although some prior work has rated model-generated UI code \cite{si2024design2code, gajos2005preference}, to our knowledge, no publicly available, large-scale dataset exists for UI design assessment.
To this end, we collected over 2.3 million UI screenshots, each paired with natural language text that includes a caption, design quality, and design defects.
Since it is prohibitively costly and time-consuming to collect enough human-annotated data to train deep learning models, the majority of our data (over 99.9\%) is synthetically generated, and a small set of human ratings is collected from designers.
We refer to our synthetically-generated dataset as \textsc{JitterWeb} and our human-rated dataset as \textsc{BetterApp}.
% We constructed training (80\%), validation (10\%), and testing (10\%) splits from this dataset, ensuring that all variations of the same URL appeared within the same split.

% The synthetic data was split into training (80\%), validation (10\%), and testing (10\%) splits, where we ensured that all variations of the same URL appeared with the same split.
% The human-rated data was split into different training (50\%), validation (10\%), and testing (10\%) splits, because a minimum numbers of samples were reserved for evaluation purposes.
% We ensured that no UI screen from the same app appeared in the same split. We describe further details on each subset's construction in the following subsections.

\subsection{Synthetic Data}

% The subset of synthetic UI data was collected by automatically crawling websites and injecting design defects using a set of scripts called \textit{jitter functions} that randomly adjust the layout and style of on-page elements.
\textsc{JitterWeb} is a synthetic dataset of 2.3 million examples created through automated web crawling, data augmentation, and captioning.
Recent research has shown that UIs on the web (\textit{e.g.,} web pages), are a useful source of data for data-driven UI modeling, due to the relative ease of applying automated crawling techniques and extracting semantic metadata from the browser \cite{wu2023webui,kumar2013webzeitgeist}.
The main idea behind our synthetic data approach was to first visit an existing web page and record its appearance (\textit{i.e.,} take a screenshot), then randomly apply several \textit{jitter functions} that intentionally degrade the design quality of the web page in different, controllable ways and record the resulting appearances.
Jitter functions are implemented as snippets of JavaScript code that, for example, add random noise to CSS attributes or swap colors in the web page's color palette.
The result of applying this process to a web page is one ``original" sample paired with several variations of itself, each with a set of known design defects.
Through this process, we are able to construct a large-scale dataset to learn the \textit{relative design quality} of UIs.
% In other words, the original screenshot is not assumed to have a ``good" design; the only assumption made is that the original screenshot has a better design than its jittered variations. 
In other words, we assume that the original UI is mostly better than the jittered one. Instances where the jittered version outperforms the original are uncommon and the noises introduced from such instances should be negligible given the scale of our dataset.

\subsubsection{Data Collection}
We followed the collection methodology of WebUI \cite{wu2023webui}, where a headless Chrome browser was used to visit thousands of websites with different simulated client devices (\textit{e.g.,} mobile phone, desktop, tablet).
It was not possible to directly re-use the publicly-released WebUI data, which consists of screenshots and extracted metadata, because our data augmentation pipeline necessitates loading the website in a browser to run the \textit{jitter functions}, which are implemented as JavaScript code.
Unlike the crawler used in WebUI, we adopted a simpler architecture that directly crawls URLs from publicly available datasets.
We crawled nearly 300,000 web pages, using URLs from the MC4 dataset provided by the Allen Institute for AI~\cite{dodge2021documenting}, which is an adaptation of the original C4 dataset~\cite{2020t5} frequently used to train large language models~\cite{le2022bloom, chowdhery2023palm, touvron2023llama, biderman2023pythia}. This dataset has undergone screening to remove explicit content~\cite{dodge2021documenting}. In addition, we excluded URLs that resulted in 404 errors.

\textsc{JitterWeb} was randomly partitioned into training (80\%), validation (10\%), and test (10\%) splits by web page URL.
We further randomly selected 201 samples from the original test split, to make it the same size as the test split from our human-rated data (\textsc{BetterApp}) for model evaluation.

% Each webpage was rendered on various simulated devices and captured using different jitter functions, leading to multiple data examples.
% The motivation behind the selection of jittering method
% connection to diffusion models
% \subsubsection{Crawling Details}

\subsubsection{Jitter Functions}
% \begin{figure}[!htb]
%     \centering
%     \includegraphics[width=\linewidth]{example-image-a}
%     \caption{A collage that illustrates example outputs from the jitter functions}
%     \label{fig:enter-label}
% \end{figure}
% \textcolor{red}{a collage that illustrates example outputs from the jitter functions}
% \textcolor{red}{why we chose these jitter functions and how they map to the design guidelines.}
\textit{Jitter functions} are JavaScript code snippets that are used to controllably introduce design defects into web pages.
To design these functions, we reviewed various guidelines on usability and design evaluation found in design textbooks \cite{shneiderman2016designing,lidwell2010universal}, online resources \cite{wong2024user,gordon2020principles}, and published literature \cite{luther2015structuring}.
While undoubtedly useful for informing application design, many of the principles described in these resources could not be assessed by looking at a single screenshot (\textit{e.g., ``error prevention,'' ``user control and freedom'')}. 
We ultimately chose the \textit{CRAP} guidelines \cite{williams2015non}, which are four general principles for UI visual design relevant to our task: contrast, repetition, alignment, and proximity.
We developed the jitter functions based on a combination of these guidelines and what is possible to programmatically adjust through JavaScript and CSS styling.
% as long as the criteria covers everything within reason.
% designers dont even have a clear textbook or rubric for evaluating designs of UI designs.
% they have different words that refer to similar or the same concepts.

Below, we describe the functions that we implemented and the CRAP principles that inspired them:
\begin{itemize}
    \item Colors
    \begin{itemize}
        \item Color Swap (contrast, repetition) - Randomly swaps the colors of elements on the web page
        \item Color Noise (contrast, repetition) - Adds numerical noise to CSS attributes for RGB values
    \end{itemize}
    \item Font
    \begin{itemize}
        \item Font Size (contrast, repetition) - Randomly swaps the font sizes of text elements in the page (\textit{e.g.,} swapping the size of subheading text with the size of body text)
        \item Text Noise (contrast, repetition) - Adds numerical noise to CSS attributes for text size
    \end{itemize}
    \item Contrast
    \begin{itemize}
        \item Text Color (contrast) - The contrast of text is decreased so that it appears closer to its container's color
        \item Background Color (contrast) - Makes the background color of containers containing text closer to the color of the text.
    \end{itemize}
    \item Spacing (alignment, proximity) - Adds numerical noise to CSS attributes for margin and padding
    \item Complexity (contrast, repetition, alignment, proximity) - Randomly removes images, text, and other element styling
    \item Layout (alignment, proximity) - Modifies CSS related to element layout such as flow (\textit{e.g.,} horizontal or vertical).
\end{itemize}

The jitter functions are composable, and when the crawler visits a web page, it chooses up to three functions via uniform random sampling to apply sequentially before taking a screenshot of the jittered UI. Figure~\ref{fig:jitters} shows an example of a web page processed by each of our jitter functions.

\subsubsection{Description Generation}
\begin{figure}[!htb]
    \centering
    \includegraphics[width=\linewidth]{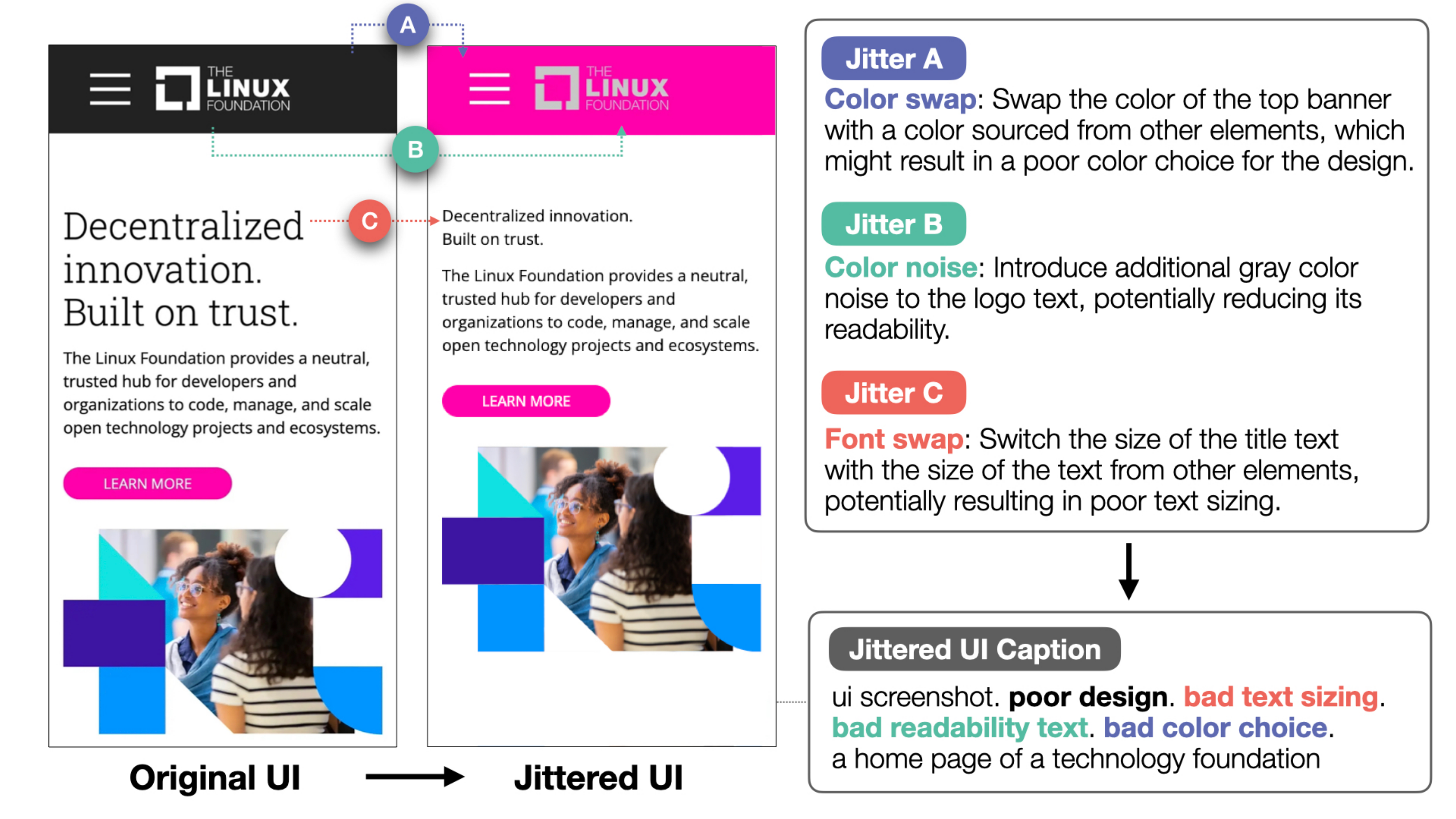}
    \caption{Our process for generating text descriptions for \textsc{JitterWeb}. Based on a set of randomly-chosen jitter functions, several design defects are introduced, \textit{e.g.,} color swap, color noise, font swap. These design defects are recorded as a part of the jittered UI's caption, which helps our model associate design defects with the UI screenshot.}
    \label{fig:jitter-caption}
    \vspace{-10pt}
\end{figure}

Each UI screenshot was associated with a natural language description that includes a caption, design quality, and a list of design defects (inferred from the applied jitter functions).
The full description is formatted by concatenating multiple components: \textit{i)} a constant prefix (``ui screenshot."), \textit{ii)} a design quality tag (``poor design" if the screen has been jittered, otherwise ``well-designed"), \textit{iii)} a list of design defects (\textit{e.g.,} if the ``text contrast'' jitter function was applied, a suggestion would be ``bad text contrast''), and \textit{iv)} a caption describing the screenshot.
Figure \ref{fig:jitter-caption} provides a visual illustration of this process.

The design-related components are inferred from the jittering process. To generate the caption, we used a set of pre-trained models to predict~\cite{lee2023pix2struct,wang2021screen2words}, then paraphrase~\cite{jiang2023mistral} a caption from the UI screenshot.
In the generation process, we specifically avoided the use of models with restrictive usage agreements or trained using data from models with restrictive usage agreements~\footnote{The terms of service of proprietary model providers such as OpenAI, Llama, and Claude prohibit using their model outputs to train other models. Therefore we avoid them and also other ``distilled'' models trained on their output.}.
Because the introduction of design defects by jitter functions may affect the accuracy of the captioning model, we generate the caption for each original UI, and then propagate the caption to all its variations.

% \textcolor{red}{Figure X} illustrates how these tags are incorporated into the full description of each screenshot.

\subsection{Human-rated Data}

While our synthetic approach to automatically generating pairs of design preferences can be efficiently scaled to millions of screenshots, it also has drawbacks.
% all of the design pairs compare against variations of the same screen and do not compare against ui screens from different apps
In the synthetic dataset, preferences are only generated between variations of the same screen, which does not reflect comparison between independent designs.
While we used established design principles to author jitter functions, they may not represent the actual distribution of design flaws across real-world apps, \textit{e.g.,} small element margins may be a very common problem ``in-the-wild" but is only represented in one of our heuristics.
Finally, a part of the creation process for the synthetic dataset involves using a pre-trained UI screenshot captioning model for caption generation. This model may produce incorrect captions that limit a downstream model's ability to understand UI design relevance.
To this end, we collected the \textsc{BetterApp} dataset using feedback from human designers.
\textsc{BetterApp} addresses the drawbacks of synthetic data by \textit{i)} comparing UI screens from different apps, \textit{ii)} collecting design defects from real apps, and \textit{iii)} using human-improved UI captions.

% \textcolor{red}{talk about synthetic examples}
As a starting point, we used an existing public dataset called VINS~\cite{bunian2021vins}, which contains screenshots of iOS apps, Android apps, design mockups, and lower-fidelity design artifacts such as wireframes.
Because it was originally used for design search and element detection applications, the VINS dataset contains screenshot images and element annotations.
For our application, we only use the screenshot images and not the lower-fidelity wireframes.
In addition to VINS data, we also included screenshots of UIs rendered by an open large-language model \cite{jiang2024mixtral} prompted to generate HTML code given natural language descriptions in our dataset.
We hypothesized that these samples would contain more variation in design quality and more design defects, which could be useful for learning design quality.

To prepare the data for our rating procedure, we applied several additional processing steps.
We first applied the same automated captioning model~\cite{lee2023pix2struct} used to construct synthetic examples to assign an initial caption to each dataset in VINS. These captions were later improved by participants.
We used a pre-trained sentence embedding model \cite{reimers2019sentence} to generate a fixed-size embedding for each screen based on its auto-generated caption. Finally, we applied the DBSCAN clustering algorithm~\cite{ester1996density} to group together screenshots with similar captions.
% captioning existing screenshots (VINS)
As a result of this process, the screenshots are collated so that screens of similar functionality can be found in the same cluster (\textit{e.g.,} all login screens). We clustered the VINS and synthetic examples separately, so clusters are only made entirely of either real-world or synthetic UIs. Designers were then asked through pairwise comparisons to assign relative rankings between UI screens in the same cluster.

\begin{figure}[!htb]
    \centering
    \includegraphics[width=\linewidth]{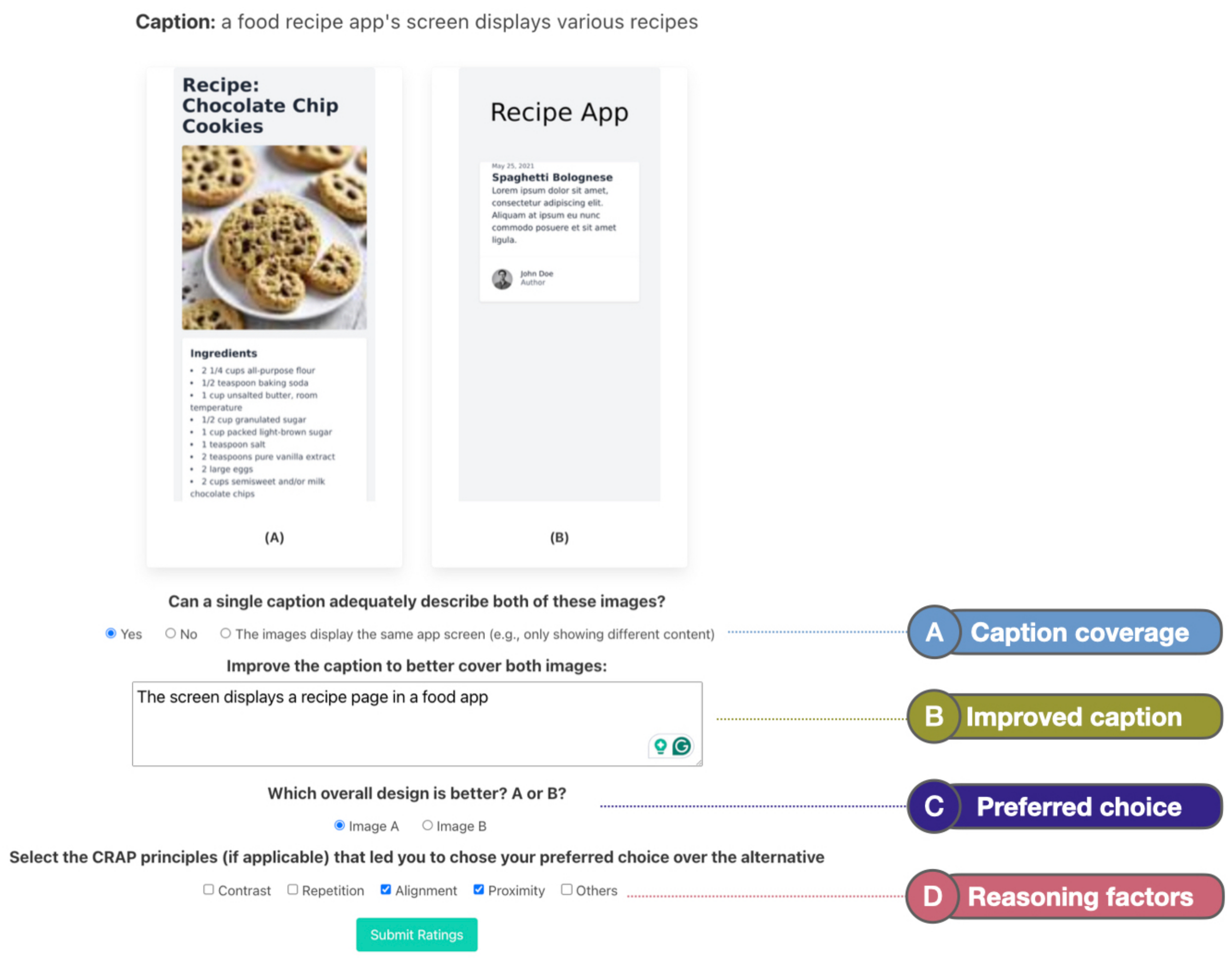}
    \caption{A screenshot of the application used for collecting human design ratings. Participants first decide whether the pair of screenshots can be described by a single caption (A). If possible, an improved caption is authored (B). Participants select one option that better matches the caption (C) and provide their reasons for doing so (D).}
    \label{fig:eval-app}
    \vspace{-10px}
\end{figure}

\subsubsection{Designer Rating Procedure}
We recruited 12 designers (ages 20-32, 11 female and 1 male) as participants at a university with varying levels of experience through word of mouth.
The participants had varying backgrounds.
Some had up to 8 years of industry experience in UI/UX design.
Others had more informal experience, but all were enrolled or had taken graduate-level courses focused on the design and implementation of UIs.
Participants spent around 1.5 hours rating UI screenshots, with the goal of reaching at least 100 screenshots.
Participants were compensated \$10 per hour (rounded up) for their time.

Participants were first asked to review an online resource that describes and provides examples of the CRAP visual design principles~\cite{kimball2013visual}.
Participant ratings were collected using a custom-built web application (Figure ~\ref{fig:eval-app}).
The start page of the application displayed instructions and recorded a visitor ID, which allowed analysis of rating consistency.

Following the start page, the web application repeatedly \textit{i)} selects a random cluster from the processed data then \textit{ii)} randomly selects two UIs from within the cluster to display.
The participant was then asked to do the following steps:
\begin{enumerate}
    \item Write a short, one-sentence caption that contains enough detail to describe both screenshots. If one of the screenshots is irrelevant (\textit{e.g.,} due to clustering error), write a caption for the first screenshot.
    \item Provide a relative ranking between the two screenshots given the options ``A is better'' or ``B is better.''
    \item Select all relevant CRAP principles that were important in determining the ranking, unless ``about the same'' was selected in the prior step.
\end{enumerate}

In total, we collected around 1200 ratings from all participants.
We ignored pairs that could not be described by a single caption, which led to 892 rating pairs. To measure inter-rater reliability (IRR), we initially had each participant evaluate the same set of 10 predetermined pairs.  Afterward, the rating pairs were distributed randomly.
% We used this initial set to compute an average Cohen's Kappa score, revealing a $\kappa$ value of 0.4.
We used this initial set to compute Krippendorff's alpha score, with $\alpha=0.37$.
% The score suggests that our raters had a fair to moderate level of agreement~\cite{mchugh2012interrater}
We discuss the factors influencing these ratings in Section 7.2, attributing the variation to the task's inherent subjectivity and variable individual preferences, such as familiarity with Android or iOS apps.

Similar to our synthetic generation approach, responses from each step are used to construct different parts of each UI screenshot's text description. The human-authored or human-refined caption from step 1 is used to improve the original auto-generated one. The relative ranking from step 2 is used to infer the correct design-quality tag, where the preferred example is assigned ``well-designed'' and the other is assigned ``poor design.'' If it was indicated that the two screenshots had the same quality, the design quality tag was omitted from the full description.
The selected principles from step 3 are used to construct a set of design defects for the non-preferred screenshot. For example, if a participant selected the \textit{contrast} principle as a reason for choosing A over B, then ``bad contrast'' is added to the generated description of B.
Note that the same screenshot can appear in more than one randomly sampled pair, which could result in conflicting descriptions \textit{e.g.,} if it was preferred in one round but not in another. Our training algorithm is robust to these collisions and over time learns to approximate a score based on the proportion of times it was preferred.

To generate \textsc{BetterApp} training (70\%), validation (10\%), and test (20\%) splits, we randomly partitioned the UI clusters, which ensured that both UI screenshots from rated pairs always occurred within the same split.
We chose the split percentages for \textsc{BetterApp} so that the size of the test set is roughly equivalent in size to other popular model benchmarks \cite{chen2021evaluating}.
The final sizes of the splits were: train (618 pairs), validation (73 pairs), and test (201 pairs).
% maybe find some mapping between the jitters and the evaluation procedure.

\section{UIClip}
% why did we choose to use CLIP as the base architecture over VLLMs?
We used the \textsc{JitterWeb} and \textsc{BetterApp} datasets to train a computational model UIClip, that assesses UI designs from screenshots.
While our datasets could be applied to train any model, such as large vision-language models~\cite{liu2023improved, bai2023qwen} that typically include the language decoder from an LLM, we adopted the CLIP architecture~\cite{radford2021learning} as it is designed to produce a numerical score, which is similar to our objective of scoring designs. 
% Maybe adding refs (if any) about LLM is not good at assigning an absolute "score"?
In addition, we also found this model to be more versatile for supporting a set of example applications (\textit{e.g.,} example retrieval and scoring) and much more efficient for training and inferencing (due to much smaller size).
There are several variations of the CLIP model, and we chose the smallest variation released by OpenAI called \textit{CLIP B/32}, which contains 151 million parameters.

\textit{CLIP B/32} is a dual-encoder transformer model (\textit{i.e.,} consisting of an image and text encoder) that accepts \textit{i)} a textual description and \textit{ii)} an image as inputs, then encodes both into a shared embedding space.
The image encoder is a vision transformer that accepts a fixed-size 224x224 image as input, splits it up into 32x32 pixel patches, and then encodes the patches into a 512-dimensional embedding.
The text encoder is a transformer that accepts text sequences of up to 77 tokens (each token roughly corresponds to a word) and also produces a 512-dimensional embedding.
The outputs of these two encoders are often used to produce a single numerical value, which is computed as the dot product of the image and text embeddings.
CLIP's dot product output can be interpreted in many ways, with a common one being the semantic similarity of the two inputs \textit{e.g.,} the text ``a dog'' and an image of a dog would produce a high score.
CLIP was trained on roughly 400 million pairs of images and text captions scraped from the internet, which it used to learn these semantic associations.
While CLIP is often successful in general image classification or association tasks, these internet crawls often lack data for more domain-specific tasks such as understanding images taken by satellites, autonomous vehicles, and medical images \cite{radford2021learning}.
As we show in our baseline evaluation, CLIP also performs poorly on UI screenshots, which are relatively rare in the model's original training data.

The purpose of our training procedure is to finetune the \textit{CLIP B/32} to \textit{i)} improve relevance scoring among UI screenshots and descriptions and \textit{ii)} incorporate design quality as a factor in the score, and \textit{iii)} associate descriptions of design defects with screenshots of UIs that contain them. We refer to our model as UIClip, since it is a descendant of CLIP that is optimized for UIs.

\subsection{Training}
We trained UIClip in four stages that incorporated different data sources and training objectives, which were designed for different use cases and tasks.
In the first training stage, which we refer to as ``pre-training," we trained UIClip using the \textsc{JitterWeb} dataset and the same training objective used in the original CLIP implementation~\cite{radford2021learning}. We found this useful for applications related to retrieval and associating UI screenshots with relevant descriptions.
In the second stage, we switched UIClip's training objective to an alternative loss function that specifically focuses on distinguishing good from bad UI designs.
These two stages are then repeated for the \textsc{BetterApp} dataset, where each stage uses model weights from the previous stage as a starting point.
During all stages of training, we adopt a pre-processing methodology similar to the one used in the original CLIP paper \cite{radford2021learning} and subsequent reproductions \cite{cherti2023reproducible}, where a random-crop strategy is used to capture different parts of UI screenshots.

\subsubsection{CLIP Pretraining Objective}

During the pre-training stage, we used the same training objective as the base CLIP model~\cite{radford2021learning}, which is described by Equation \ref{eq:training_loss}.
\begin{equation}
    \mathcal{L}_{CLIP} = {\displaystyle -\sum _{i}\ln {\frac {e^{v_{i}\cdot w_{i}}}{\sum _{j}e^{v_{i}\cdot w_{j}}}}-\sum _{j}\ln {\frac {e^{v_{j}\cdot w_{j}}}{\sum _{i}e^{v_{i}\cdot w_{j}}}}}
    \label{eq:training_loss}
\end{equation}
Where, $w_{i}$ refers to the $i$-th text embedding in the batch and $v_{j}$ refers to the $j$-th image embedding in the batch.

To give a high-level overview of the process, this training objective involves repeatedly sampling a \textit{minibatch} of $N$ examples from the training dataset, where each example consists of an image (UI screenshot) and a textual description (caption with a design quality tag and applied jitters).
The model generates embeddings for all text $w_{1...N}$ and images $v_{1...N}$ in the minibatch, then computes an $NxN$ similarity matrix between all combinations of images and text.
The objective then computes the cross entropy loss to match each image with its original text description, and vice versa.
The intuition behind this process is that the representations of corresponding images and text will gradually become more similar in the shared embedding space, while mismatched pairs will be pushed apart.
In the case of UIClip, screenshots will be matched to textual descriptions containing the appropriate design quality tag, design suggestions, and caption.

\subsubsection{Pairwise Contrastive Objective}
A drawback of the standard CLIP objective is that the minibatches used to compute its loss are randomly sampled from the entire training dataset.
Because the size of a minibatch is much smaller than the size of the entire training dataset, there is very low chance that a minibatch will contain examples of closely-related UI screenshots \textit{e.g.,} both a jittered and non-jittered version of a webpage.
We hypothesized that this would make it more difficult for the model to learn relationships between these related UIs, which is necessary for assessing the relative quality of related designs.
Therefore, we modified the training objective to explicitly compare pairs of related UI screens.
Our method is similar to previous methods for pairwise contrastive learning~\cite{hadsell2006dimensionality}, but we use a cross-entropy loss, which is more compatible with the pre-training objective, instead of the margin-based one.

This training objective, shown in Equation \ref{eq:pairwise_loss}, trains the model so that the embedding of the preferred screenshot has a higher dot product with a text description indicating good design (\textit{i.e.,} a design quality tag of ``well-designed'') than the embedding of the non-preferred screenshot.
\begin{equation}
    \mathcal{L}_{pair} = {\displaystyle -\ln {\frac {e^{v^{+}\cdot w^{+}}}{e^{v^{+}\cdot w^{+}} + e^{v^{-}\cdot w^{+}}}}}
    \label{eq:pairwise_loss}
\end{equation}
Where $v^{+}$ refers to the embedding of the preferred screenshot, $v^{-}$ is the embedding of the non-preferred screenshot, and $w^{+}$ is the embedding of the text description.

\subsection{Inference}
\label{sec:inference}
% \textcolor{red}{figure visualizing the inference modes}

\subsubsection{Preprocessing}
% \textcolor{red}{cropping vs padding. sliding window algorithm,}
% \textcolor{red}{talk about how the image is scaled for the input}
CLIP has a fixed image input size of 224x224 pixels, which presents challenges for encoding UI screenshots during inference given that many mobile apps are designed with high height-to-width aspect ratios and dimensions can vary significantly between UIs captured on different devices.
Naive pre-processing methods such as image scaling or image cropping can result in significant distortion or exclude important information.
One way to address this is to make architectural changes to the model, using similar strategies to previous work \cite{lee2023pix2struct}. % Pix2struct
We adopt a simpler strategy to handle variable image sizes using a sliding window strategy. % cite paper about using CLIP handling video 
The input screenshot is first resized so that its smaller dimension is equal to 224 pixels.
A 224x224 window slides across the larger dimension of size $d$ where the number of evenly-spaced steps is equal to $\lfloor \frac{d}{224} \rfloor + 1$, so that the entire image is covered with the minimal amount of overlapped area.
The image encoder is used to compute an embedding for each window of the screenshot, and then all embeddings are averaged together.

\subsubsection{UIClip Score}
The UIClip score represents a combination of the relevance of the text description and the UI screenshot and the design quality of the UI screenshot.
Computing the UIClip score requires \textit{i)} a screenshot of the UI to be evaluated and \textit{ii)} a user-provided caption describing the intended purpose of the UI.
A full textual description is constructed by pre-pending a prefix ``ui screenshot. well-designed. " to the user-provided caption.
The resulting score between the encoded screenshot and the encoded full description represents a score that describes how well the screenshot adheres to a ``well-designed" UI with the target caption.

\subsubsection{Design Suggestions}
% inference of energy-based models is NP-hard and must be approximated.
% we should probably add a sentinel token to the captions in the dataset
% bert has a mouth and it must speak
% Because the UIClip is non-Markovian \footnote{Inference of Markovian lattices can be efficiently computed using the Viterbi algorithm [x].}, finding the set of tags that maximizes the model score is a computationally complex problem that requires iterating through every possible set of tags.
During training, UIClip learns to associate screenshots with with natural language descriptions containing design defects that are potentially contained within them.
However, because UIClip doesn't contain a decoder network, it cannot directly generate text like other auto-regressive transformers \cite{radford2018improving}.

Instead, we develop an alternative approach that uses UIClip to detect possible design defects in an input screenshot, then surfaces them as warnings to the user to fix.
For each possible defect, a natural language description is constructed by pre-pending the corresponding prefix to the caption, \textit{e.g.,} ``ui screenshot. poor design. bad text sizing. login screen.''
We consider all design defects introduced by our jitter function (\textit{e.g.,} ``bad text sizing'') and the four CRAP principles (\textit{e.g.,} ``bad alignment'').
We computed the similarity score between the input image and these text descriptions that corresponded to design defects.
To determine the design defects that are surfaced, we dynamically compute a threshold. The threshold is computed as the image's similarity score with a caption without any defect tags, \textit{e.g.,} ``ui screenshot. poor design. login screen.''
% The defect used to generate the description with the highest score is selected as a suggestion.
% probably update this to a threshold that is computed via the "distribution of softmax scores"
Design suggestions can also be limited to a smaller number of categories (\textit{e.g.,} only the four CRAP principles) through pre-defined mappings.
For example, since the color noise jitter could affect both contrast and repetition, we map ``bad color choice'' to warnings for these classes.

\begin{figure*}[!t]
  \centering
  \includegraphics[width=1.0\textwidth]{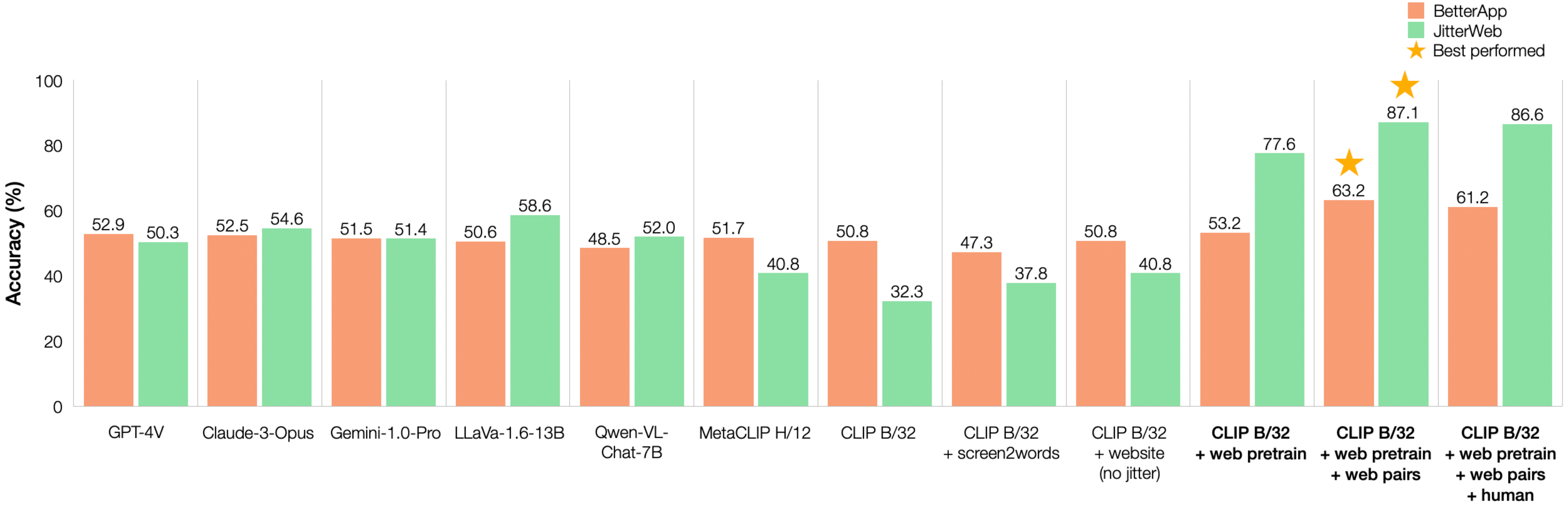}
  \vspace{-20px}
  \caption{Model performance on \textit{design choice} prediction, which involves identifying the preferred UI screenshot from a pair. UIClip models (with bold font) perform the best on held-out human-rated pairs from \textsc{BetterApp} and synthetically-generated pairs from \textsc{JitterWeb}. Most baselines perform poorly, around the level of random chance.}\label{fig:chart_design_choice}
  \Description{}
\end{figure*}

\section{Evaluation}
The purpose of our evaluation is to quantify multiple aspects of UIClip's design assessment capabilities and to compare its performance against several state-of-the-art baseline models and ablation conditions.
We focus on tasks that correspond to three use-cases: \textit{i)} design quality assessment, \textit{ii)} design suggestion generation, and \textit{iii)} design relevance.
% \textcolor{red}{F1=X} and is more effective than other models in language-driven example retrieval
In all three tasks, UIClip outperforms baseline models that are often several orders of magnitude larger.
\subsection{Procedure}
We conducted a quantitative evaluation that measured model performance using held-out examples from our datasets.
% add a table to show parameter sizes of different vision and language encoder and model 

\subsubsection{Baselines}
We chose several baselines that consist of different types of multimodal machine-learning models.
Originally, we planned to include AIM~\cite{oulasvirta2018aalto}, which is a software package for computing various metrics for UIs.
However, there is no definitive way to convert these metrics into design ratings, so we excluded it as a baseline.
Therefore, we limit our analysis to the machine-learned models described below.

% remove the bullet points from this itemize
\begin{itemize}
    \item \textbf{Proprietary Large Vision-Language Models (only accessible via APIs)} 
    \begin{itemize}
        \item \textit{OpenAI GPT-4V} - GPT-4V is a model developed by OpenAI that has been shown to excel at a variety of tasks \cite{achiam2023gpt}.
        \item \textit{Anthropic Claude-3-Opus} - Claude-3-Opus is a model that was introduced by Anthropic at March 2024. It is the largest and most powerful variant among the three Claude-3 models~\cite{claude3}.
        \item \textit{Google Gemini-1.0-Pro} - Google Gemini-1.0-Pro (Vision) is a model that was introduced by Google in December 2023~\cite{team2023gemini}. It's the most powerful publicly available model among the three Gemini-1.0 models (Gemini-1.0-Ultra was announced but not publicly available at the time we performed this benchmarking).
    \end{itemize}
    \item \textbf{Open-source Large Vision-Language Models (model weights are publicly available)}
    \begin{itemize}
        \item \textit{LLaVA-1.6-13B} - LLaVA~\cite{liu2024llavanext} is a model that was trained using a combination of training examples generated by proprietary large language and vision-language models as well as publicly available visual reasoning and image caption datasets. We used the 13B model (with ViT-L/14~\cite{zhai2022scaling} as the vision encoder and Vicuna-13B~\cite{vicuna2023} as the language decoder) as one of our baselines, which is the largest model that we could fit on our GPU hardware.
        \item \textit{Qwen-VL-Chat-7B} - Qwen-VL-Chat~\cite{bai2023qwen} is similar to LLaVA, but it was trained using an alternative pre-training objective and datasets. This model (with ViT-bigG~\cite{gadre2024datacomp} as the vision encoder and Qwen-7B~\cite{bai2023qwen} as the language decoder) is notable because its pre-training data contained images of documents, which we hypothesized could be relevant for understanding UIs as well.
    \end{itemize}
    \item \textbf{CLIP Models}
    \begin{itemize}
        \item \textit{CLIP B/32} - We used the unmodified \textit{CLIP B/32} model, which was trained by OpenAI on 400M image-text pairs known as the WebImageText dataset.
        \item \textit{MetaCLIP H/12} - Recent research has focused on improving the performance of CLIP models by scaling model size~\cite{cherti2023reproducible} and curating larger and higher-quality multi-modal training datasets~\cite{gadre2024datacomp}. \textit{MetaCLIP H/12} is a recent model to achieve state-of-the-art performance for CLIP-like models. It is roughly 6 times larger than \textit{CLIP B/32} and was trained on roughly 6 times more data~\cite{xu2023demystifying}.
    \end{itemize}

    \item \textbf{CLIP Models with Alternative Data}
    \begin{itemize}

        \item \textit{CLIP B/32 + Screen2Words} - We trained \textit{CLIP B/32} on the largest (to our knowledge) publicly-released dataset of UI screenshots paired with human-authored natural language captions using the default CLIP training objective.
        \item \textit{CLIP B/32 + non-jittered websites} - We trained \textit{CLIP B/32} on only non-jittered websites from \textsc{JitterWeb} using the default CLIP training objective.

    \end{itemize}
    \item \textbf{UIClip}
    \begin{itemize}
        \item \textit{CLIP B/32 + jittered websites} - We trained \textit{CLIP B/32} on all data from \textsc{JitterWeb} using the default CLIP training objective.
        \item \textit{CLIP B/32 + jittered websites + web pairs} - We trained \textit{CLIP B/32} on all data from \textsc{JitterWeb} using both the default CLIP objective and the paired contrastive objective.
        \item \textit{CLIP B/32 + jittered websites + web pairs + human pairs} - This model consists of \textit{CLIP B/32} trained on \textsc{JitterWeb} and \textsc{BetterApp} using both the default CLIP objective and the paired contrastive objective.
    \end{itemize}
    
\end{itemize}

\subsubsection{Model Inference}
LVLM models rely on a natural-language prompt to instruct them on how to process the image input.
We constructed a prompt that asked the model to use the CRAP principles to choose the better design of two UI screenshots and provide the most relevant CRAP principles for its decision.
We provided the model with the same short description of the CRAP principles we gave our designers who rated the \textsc{BetterApp} dataset.
Since some models could only accept one image input, we concatenated two UI screenshots side by side into a single image.
In preliminary tests, we verified that all models could distinguish the inputs by asking them to describe the left and right screenshots of the input image.

We iterated through several versions of prompts which included well-known strategies for eliciting step-by-step reasoning \cite{kojima2022large}.
We chose the best natural language prompt from our tests and used it for all models. The format of the prompt is provided in the appendix (We also included an example of GPT-4V's output when accessed through the web interface in Figure \ref{fig:gpt-example}, with a slightly modified prompt that allowed it to provide additional reasoning). We used the default parameters (e.g., temperature and top-p) for all the LVLMs we compared.
% YH: seems kinda weird to put it here as "inference" as this is not what we did for evaluation but just one example we tried to show GPT4-V did get wrong answers (?)
UIClip and other CLIP-based models used the inference strategies described in Section \ref{sec:inference}.

% \subsubsection{Metrics}
 % reference CLIPScore

\begin{figure*}[!t]
  \centering
  \includegraphics[width=1.0\textwidth]{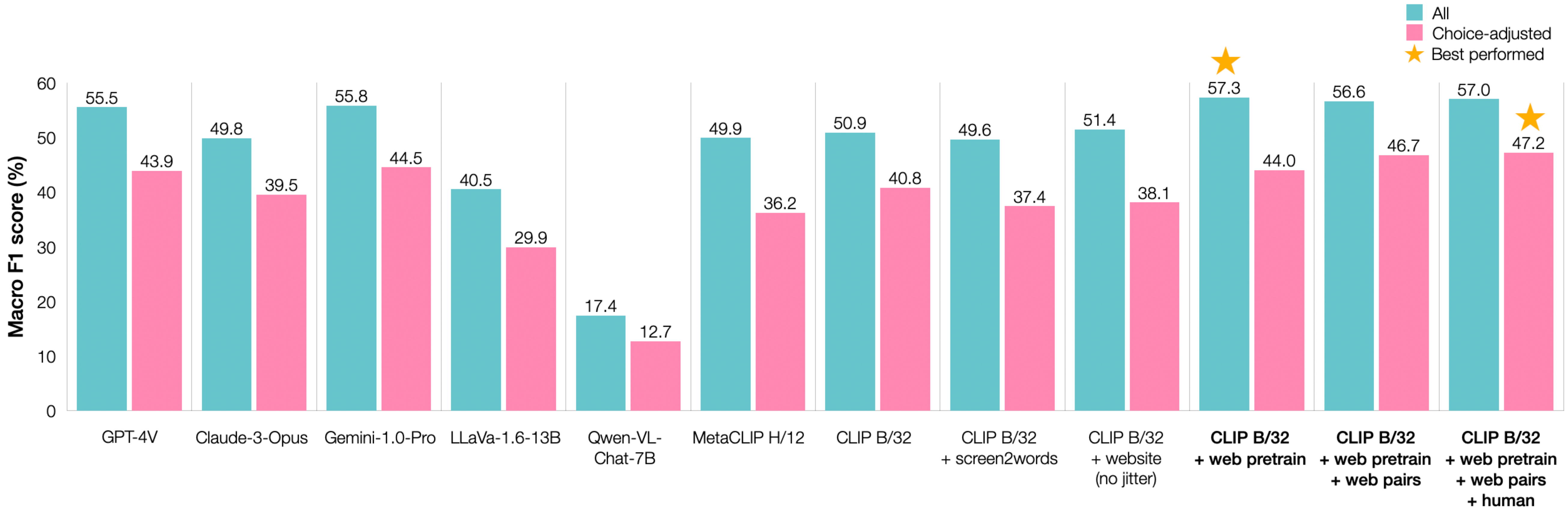}
  \vspace{-20px}
  \caption{Model performance on \textit{design suggestion} prediction, which involves generating design suggestions for a UI based on detected design flaws. We used the macro-averaged F1 score to measure performance across four CRAP principles. In addition, we introduce a choice-adjusted metric that ignores generated suggestions if they led to the incorrect choice. Using both metrics, UIClip models (with bold font) perform the best on held-out pairs from \textsc{BetterApp} and \textsc{JitterWeb}.}\label{fig:chart_design_suggestion}
  \Description{}
\end{figure*}

\subsection{Results}

% \subsubsection{Relevance}
% \subsubsection{Quality}
% \subsubsection{Recommendations}
% We evaluated the accuracy of model-generated design recommendations
We focused on evaluating three aspects of design assessment: \textit{i)} UI design quality assessment, \textit{ii)} design suggestion generation, and \textit{iii)} design relevance.

\subsubsection{Design Quality}
We evaluated a model's \textit{design quality} assessment by measuring its accuracy in identifying the ``preferred'' UI from an example pair.
The results of our experiments are shown in Figure \ref{fig:chart_design_choice}.
% In total, there were \textcolor{red}{X instances.}

% The results of the design quality evaluation are shown in \textcolor{red}{Figure X}.
In general, design quality assessment is a difficult task for all tested models, especially when evaluating human-rated pairs of real app screens.
Our results are shown in Figure \ref{fig:chart_design_suggestion}.
For both \textsc{BetterApp} and \textsc{JitterWeb}, the UIClip variant trained with web pairs performed the best, with an average overall accuracy of 75.12\%. 
In particular, we see large improvements in detecting design defects in web pages (87.1\%), which is what the majority of the training process and data focused on.

These improvements are notable because \textit{CLIP B/32}, which was the base model of all UIClip variants, performed the worst out of all baselines. \textit{CLIP B/32} performed especially poorly for jittered websites, where a further analysis revealed it erroneously associated certain types of jitters (\textit{e.g.,} dark, unreadable backgrounds) with better design. This suggests that our training procedure and data are effective for learning design quality, especially when compared to other publicly available sources of captioned UIs (\textit{e.g.,} Screen2Words) or general-purpose multi-modal data (\textit{e.g.,} \textit{MetaCLIP H/12}).

Incorporating human ratings appeared to lead to slightly degraded performance (overall accuracy of 73.88\%), possibly due to insufficient data.
As noted previously, the UIClip with CLIP pre-training objective alone was less effective at improving design quality assessment capabilities because paired UI examples are often not found in randomly sampled minibatches. Nevertheless, it had the third-highest overall accuracy of 65.42\%.

Despite their much larger size, all LVLMs performed very poorly on design quality assessment, often around the level of random guessing. Interestingly, GPT-4V (overall 51.58\%) refused to provide a response for around 10\% of examples, stating ``I'm sorry, I can't help you with that.''
In this regard, open models such as LLaVA performed better than GPT-4V, even though LLaVA was trained by distilling GPT-4V output. However, since LLaVA was not trained to refuse requests \cite{liu2023improved}, it ended up with a higher overall performance for design choice prediction (with 54.59\% prediction accuracy).

\begin{figure}[!htb]
    \centering
    \includegraphics[width=\linewidth]{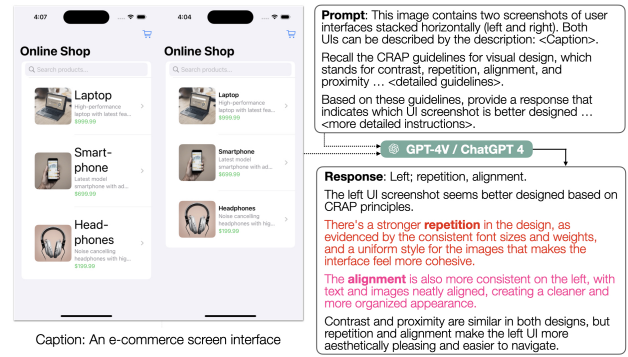}
    \caption{An example of incorrect design assessment and reasoning from GPT-4V. We asked GPT-4V to compare two screenshots. We fed in the same prompt used in our baseline evaluation, except in this case, we allowed GPT-4V to list its reasoning. The screenshot on the right is the original application screenshot. The screenshot left is a variation of the same interface with a design defect where the text is too large and overflows, affecting readability and coherence. GPT-4V erroneously states the left screenshot is better designed and provides irrelevant and incorrect reasoning.}
    \label{fig:gpt-example}
    \vspace{-10px}
\end{figure}

\subsubsection{Design Suggestions}
We evaluated all models' design suggestion capabilities by comparing the model-generated output to the CRAP principles selected by designers when rating UI quality in \textsc{BetterApp}.
There were four possible CRAP principles that could have been chosen for each comparison, which we formulate as a multi-label classification problem with four labels.
Since designers were allowed to omit reasoning for comparisons, we ignored comparisons where none of the CRAP principles were selected.

Again, design suggestion was a challenging task for all tested models.
Some LVLM baselines listed all four CRAP principles for almost every single example, despite being prompted to only choose the most relevant principles.
This appears to be consistent with prior work on using LLMs for heuristic evaluation \cite{duan2024generating}, where similar models often provided a large number of irrelevant design suggestions.

In our case, this phenomena led to artificially high recall for models such as Gemini (87.11\% recall) and GPT-4V (84.57\% recall).
Thus, we introduced a \textit{choice-adjusted} F1 metric that ignored models' design suggestions if it led to choosing the wrong preferred UI, \textit{i.e.,} right reasoning but wrong answer.
This lowered all models' recall to more realistic levels, \textit{e.g.,} Gemini's recall was lowered to 49.17\% and GPT-4V was lowered to 46.58\%.
Some open LVLM baselines, such as Qwen-VL-Chat, also had trouble following our prompt and often ignored instructions that asked them to provide reasoning for their answers.

Under both methods of calculation, UIClip variants had the best performance, with the web pre-trained variant performing the best when all examples were considered and the full UIClip variant performing the best when adjusted for choice accuracy.

CLIP variants that were trained on other sources of data did not achieve high performance, since their training data did not include information about present design defects that would help them make suggestions.

\subsubsection{Design Relevance}
% \begin{figure}[!htb]
%     \centering
%     \includegraphics[width=\linewidth]{example-image-a}
%     \caption{Ground truth retrieval results. bar graph}
%     \label{fig:enter-label}
% \end{figure}

Finally, we also evaluated a model's ability to compute UI relevance, which is useful for assessing designs and for various applications that require example retrieval~\cite{bunian2021vins,huang2019swire}.

To measure UI relevance, we adopted a metric commonly used in information retrieval known as mean reciprocal rank (MRR).
An embedding is computed for the preferred screenshot in \textsc{BetterApp} and \textsc{JitterWeb}.
For each description in the evaluation set, we appended the prefix ``ui screenshot. well designed. " and computed its text embedding.
The text embedding is used the calculate similarity scores with all screenshots, which is used to sort them in descending order (\textit{i.e.,} highest similarity first).
The rank of the first element with the same description was recorded.
Since a lower rank is desirable (indicating higher similarity with the corresponding image), MRR (higher is better) is computed as the average of all \textit{reciprocal} ranks.
A higher MRR indicates better retrieval performance.
Because there is no straightforward way to generate rankings from LVLMs, we only evaluate models based on the CLIP architecture.

% The results of the UI relevance evaluation are shown in \textcolor{red}{Figure X}.
The results of our retrieval evaluation are shown in Table \ref{tab:design-relevance}.
The variant of UIClip pretrained on \textsc{JitterWeb} using the default CLIP objective achieves the highest MRR score for both \textsc{BetterApp} (0.3851) and \textsc{JitterWeb} (0.4085).
UIClip variants trained using pairwise loss were the \textit{worst} performers, with MRRs lower than the original \textit{CLIP B/32} base model, because the objective focuses on the design-comparison task.
In our discussion, we provide more detailed reasoning for this phenomenon.

Nevertheless, our evaluation still shows that our datasets are useful for learning design relevance, especially when training objectives are closely aligned.
For example, applying the CLIP objective to \textsc{JitterWeb} is much more effective than alternate data sources and nearly doubles ($0.2000 \rightarrow 0.3968$) the overall retrieval performance of the base \textit{CLIP B/32} model.

\begin{table}[]
% \tiny
% \caption{Model Performance on UI Retrieval for both BetterApp and JitterWeb datasets}~\label{tab:design-relevance}
% \vspace{-0.5cm}
\caption{Model Performance on UI Retrieval for both \textsc{BetterApp} and \textsc{JitterWeb} datasets. The variant of UIClip trained with the CLIP pretraining objective (\textit{i.e.,} without paired data) performed the best, while other variants of UIClip had poor performance, due to the pairwise objective's high prioritization of design choice accuracy.}~\label{tab:design-relevance}
\resizebox{3.33in}{!}{%
\begin{tabular}{@{}lll@{}}
\toprule
Model                        & MRR (BetterApp) & MRR (JitterWeb) \\ \midrule
MetaCLIP H/12                & 0.2722          & 0.2350          \\
CLIP B/32                    & 0.2534          & 0.1466          \\
CLIP B/32 + Screen2Words     & 0.2938          & 0.1130          \\
CLIP B/32 + Web & 0.3467          & 0.3766          \\
CLIP B/32 + Jit. Web     & \textbf{0.3851}          & \textbf{0.4085}          \\
CLIP B/32 + Jit. Web + Web Pairs    & 0.0962          & 0.0924          \\
CLIP B/32 + Jit. Web + Web Pairs + Human   & 0.1096          & 0.1214          \\
\bottomrule
\end{tabular}
}

\vspace{-0.5cm}
\end{table}

\section{Example Applications}
Based on the three capabilities of UIClip that we evaluated, we present a suite of example applications that illustrate how common user-facing design tools can be enhanced with our model.

\subsection{Improving UI Code Generation}
\label{sec:uigeneration}
\begin{figure}[!htb]
    \centering
    \includegraphics[width=\linewidth]{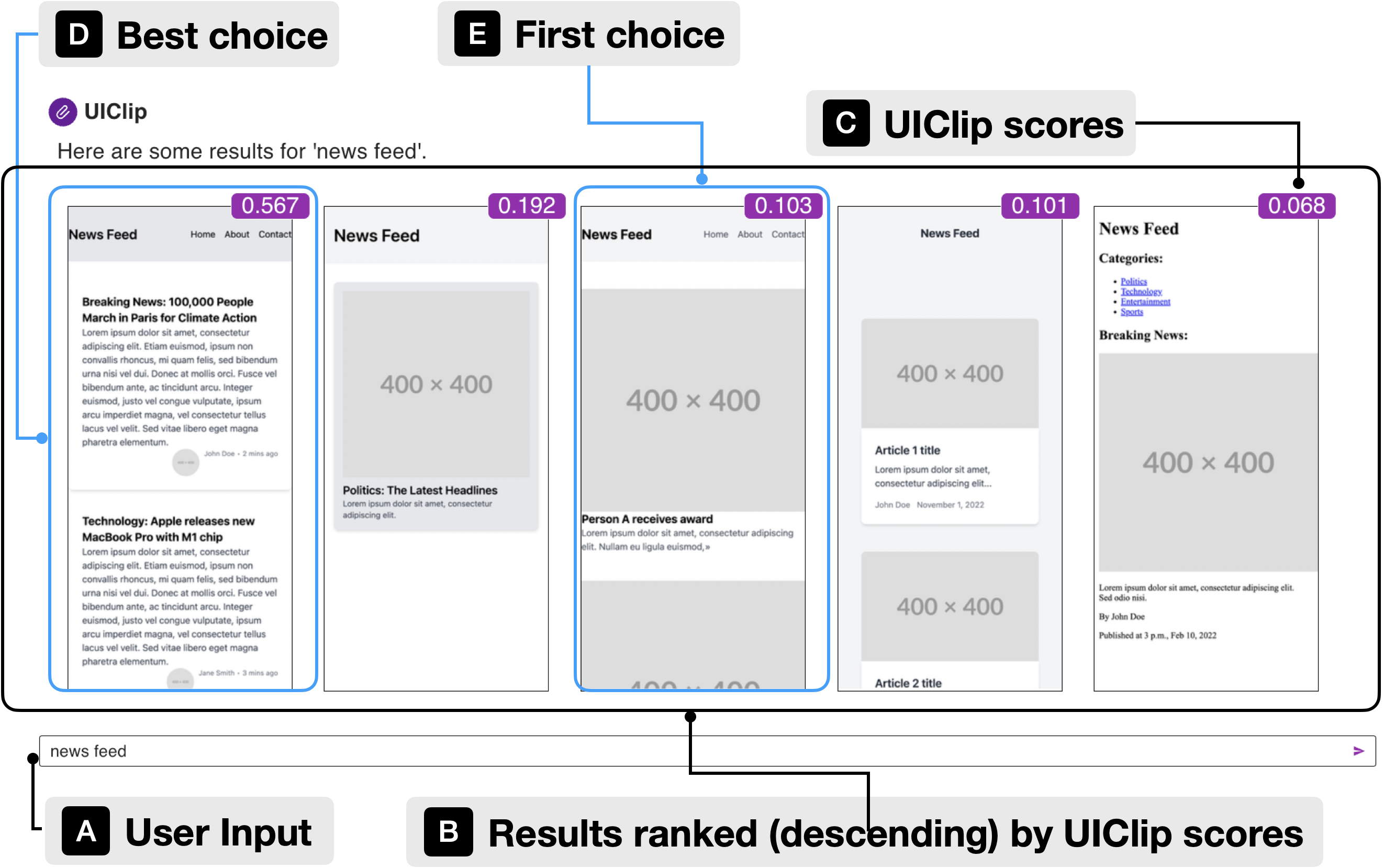}
    \caption{UI Generation Example Application. UIClip is used to rank rendered UI code provided by an external LLM (GPT-3.5). The user can describe a UI (A) and compares different LLM-generated results (B). We ranked these options based on UIClip's scores and displayed them alongside the rendered screenshot (C). Note that UIClip ranked the first screenshot on the left as the highest quality (D), while we received the third screenshot from the left as the first choice output from our LLM (E).}
    \label{fig:example_codegen}
\end{figure}

% \textcolor{red}{talk about the importance of cheap (monetary and speed) inference for applications}
We built a web application that allows users to generate rendered UI screenshots from a natural language description of a UI. 
% cite uicoder, huggingface websight demo, sebastian bubeck's paper
To use the interface (Figure \ref{fig:example_codegen}), users enter their descriptions in a textbox, which is formulated in a prompt. The prompt is fed into an external LLM (OpenAI GPT-3.5), which generates web code (HTML/CSS) using the description. We sampled $n=5$ different outputs and rendered each into a screenshot using the script that programmatically controlled a browser. If the web code referenced external images, we replaced them with a placeholder image to render. Screenshots were fed into UIClip and were scored against the input prompt. The screenshot of the results ranked in descending score order is returned to the user.

This is a simple example of how UIClip could be used to improve the output of generative models, most similar to existing ``best-of-n sampling'' approaches.
The method can also be incorporated into the additional vision checkup~\cite{sharma2024vision} and used for feedback in self-improving generative model outputs~\cite{madaan2024self}.
Our technique is simple to implement and does not require access to the underlying model's weights; however, it is computationally expensive during inference because multiple candidate solutions must be generated.
If model weights are available, this process could be further improved by applying UIClip's filtering during the training process of generative models, or if UIClip was used as a reward model in reinforcement learning fine-tuning approaches~\cite{ouyang2022training, gulcehre2023reinforced}. 
% cite ouyang rlhf, rest, uicoder? 
% dalle was trained like this using clip 
We leave these investigations to future work.

\subsection{UI Design Tips}
\begin{figure}[!htb]
    \centering
    \includegraphics[width=\linewidth]{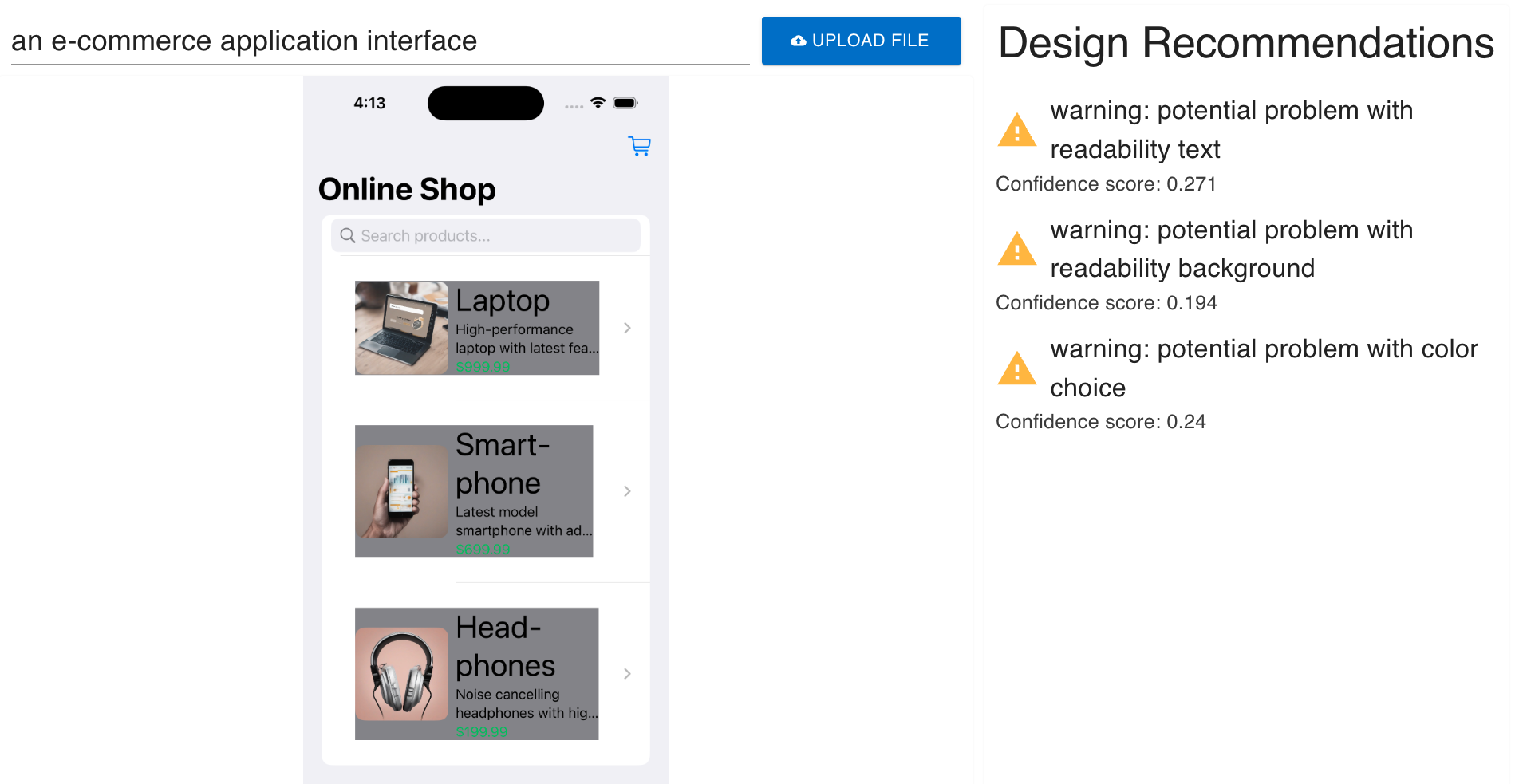}
    \caption{UI Design Tips Example Application. We use our design suggestion generation algorithm to generate design suggestions from a user-provided description and user-uploaded UI screenshot. This example shows suggestions to improve the readability of text and color choice for the application.}
    \label{fig:example_recommend}
\end{figure}
We built a tool that allows users to upload screenshots of UI designs to generate design tips using our model's design suggestion capabilities.
We developed a web application (Figure \ref{fig:example_recommend}) that allows users to upload a screenshot of an app or UI design, and our system generates tips that are surfaced to the user, although a similar idea could be better integrated into, for example, UI design applications for improved ease-of-use. One limitation of our current application is that it might suggest improving text contrast, but it is unable to provide additional information for which part of the UI led to the recommendation. This is due to our problem formulation that pairs descriptions with entire screenshots and doesn't contain location information. Future improvements can help address this by collecting the relevant data and incorporating that into text descriptions, or by sliding a smaller window across the UI screenshot and associating generated design suggestions to the location of the window.
We leave these additional features to future work.

\subsection{UI Example Retrieval}
\begin{figure}[!htb]
    \centering
    \includegraphics[width=0.75\linewidth]{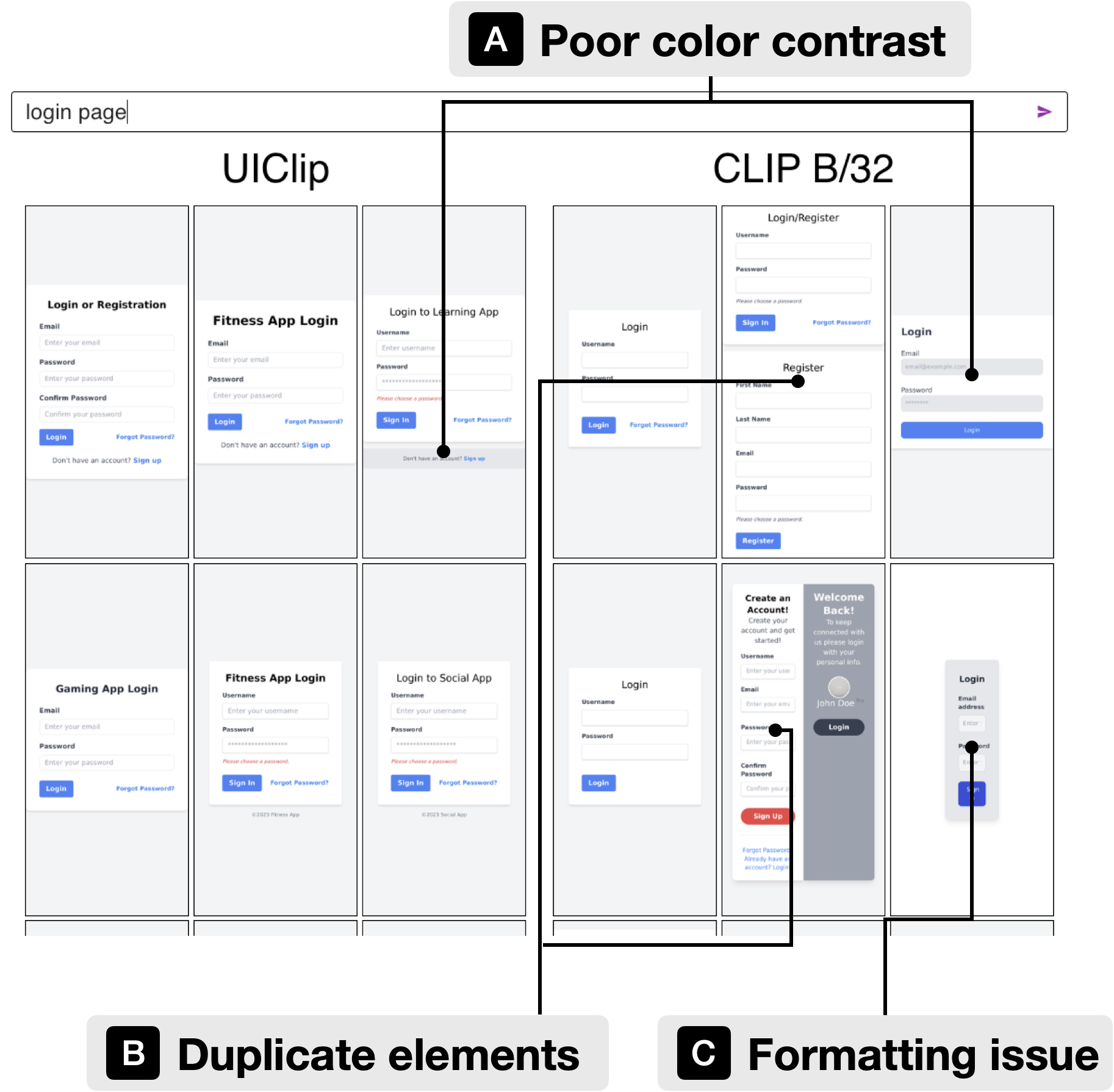}
    \caption{UI Example Retrieval Example Application. In this use case, the designer searches for examples of login screens queried from a set of LLM-generated UIs, many of which have design flaws. While both UIClip and \textit{CLIP B/32} gives a diverse range of applications, we see that there are more design flaws present from \textit{CLIP B/32}. Some screens exhibit poor color contrast (A) which may imply that the component is disabled, duplicate elements (B) that can confuse end users, and overall poor formatting and overflow layouts (C).}
    \label{fig:example_search}
\end{figure}
% VINS, screen parsing, galleryDC, RICO, RICO follow up by thomas liu.
UI design search has been explored by many prior works \cite{kumar2013webzeitgeist,deka2017rico,bunian2021vins}, and it has the potential to accelerate the design process by providing relevant examples that designers use to seek inspiration during early phases of the design process. % moodboarding
% \textcolor{red}{talk about UIClip's embedding model facilitates this, as opposed to GPT-4}
Existing systems built for UI example retrieval have focused on querying and indexing UI screenshots by their layout (\textit{e.g.,} to support finding designs similar to an exemplar) or captions (\textit{e.g.,} to support natural-language or free-form search).
Since UIClip contains both an image and text encoder, it is possible to support both of these use cases, although our example application focuses on handling text-based queries.
Our application uses a similar procedure to our UI relevance evaluation, where model-computed embeddings are used to retrieve and sort screenshots based on the user's query.
UIClip's score can take into account both the relevance and quality of retrieved examples, and we incorporate a \textit{negative prompt} that biases the query vector away from simple or ambiguous designs \cite{sanchez2023stay}.% Stay on topic with Classifier-Free Guidance

We built a web application that contains a search box where the user enters their query. Figure \ref{fig:example_search} shows examples of screens retrieved for a set of queries indexed by UIClip and the vanilla CLIP model. 
% equation for vector similarity matching

% \section{Discussion}

% humans look for consistency but might be hard to automatically evaluate that
% design recommendations would be more valuable if it was specific enough so that it pointed out parts of the screen that could be improved and provided more operationalizable feedback in general.
% \subsection{Limitations \& Future Work}
% trained on the web
% focus of evaluation and human-annotated datasets on mobile app interfaces
% would this work for all UIs

\section{Discussion}
Our experiments and example applications show that UIClip's design assessment capabilities can improve many machine-assisted design tools. In this section, we discuss UIClip's implications, limitations, and directions for future work.

\subsection{Data-driven Learning of UI Design}

% More recently, similar frameworks have been applied to understand how users interact with
% gajos's paper focuses on an individual person

Our paper introduces techniques for machine-learning a \textit{generalized} scoring function (\textit{c.f.} personalized functions \cite{gajos2005preference}) that quantifies aspects UI design.
% UIClip could be used to support human designers (shown by our example applications), improve automated UI generation frameworks \cite{gajos2004supple}, and serve as a ``reward model" to guide the training of larger models \cite{ouyang2022training}.
We discuss our work's data and algorithm contributions.

We hypothesized that a large volume of data (millions of examples) is important for effectively learning to assess designs, similar to how seasoned human designers develop their intuition through years of experience.
This hypothesis was largely supported by our experimental results.
We showed the substantial benefits of training on our large-scale dataset of UI designs, leading to improvements over alternate datasets (\textit{e.g.,} Screen2Words~\cite{wang2021screen2words}) that more \textit{human-authored} descriptions but fewer overall samples.
When we incorporated our own human-rated \textsc{BetterApp} dataset, we found that in most cases, it did not result in substantial changes, most likely due to insufficient data volume.

% design volume
% maybe this can be rolled into the previous one
% why human pairs sometimes does worse
% talk about screen2words vs websites jittered
% computational rationality by antti
% science of the artificial
% hard to assign absolute value, but easier to learn relative value
% comparisons rather than rubric
% talk about why pairwise vs rating (we did this a little bit before but re-stating)
% relative vs absolute quality
% bradley-terry model's equivalence between preference scoring and pairwise ratings and Elo

At the same time, dataset size alone is not enough to ensure good design assessment performance.
For example, \textit{MetaCLIP H/12} was trained on a total of 2.5 billion pairs.
Our paper introduces training objectives targeted at different aspects of design quality.
We employed two objectives for training UIClip, a batch-wise contrastive objective (\textit{i.e.,} CLIP's pretraining objective) and a pairwise contrastive objective, designed specifically for quality comparisons.
% Models trained on the same data with loss functions excelled at different tasks: the pre-training objective led to best performance in retrieval tasks while pairwise objective led to the best performance in design-comparison tasks.
Based on our results, the pairwise objective represents a tradeoff where it achieves higher focus on design-comparison tasks by focusing on pairs of relevant screens but incurs a penalty on retrieval-related tasks, since it is not trained to distinguish irrelevant examples from a diverse minibatch.
% Because the pairwise loss is designed specifically for the \textit{design quality} task, it only trains the model to compare the quality of two UIs (with the same caption) at once, which improves comparison accuracy but degrades model performance in distinguishing caption relevance. 
% In contrast, under the default CLIP objective, the model learns to associate the most likely image and text within a diverse minibatch sampled from the training dataset. However, it is less likely to encounter pairs of related screens that it can use to learn relative design quality.
We leave further investigation of how to optimally combine these two training objectives to future work; although given the relatively small size of our model, we believe it would be feasible to use different variations for application-specific scenarios.

\vspace{-2.5px}

\subsection{Formulating UI Design Quality}
UIClip's current model architecture is designed around the assumption that design quality can be represented by a numerical score.
However, there are many nuances that cannot be captured by this formulation.

Within the context of our collected data, we distinguish between assessing UIs for \textit{design defects} and understanding more subtle \textit{design preferences}.
\textsc{JitterWeb} was constructed by introducing ``jitters'' into web pages, that intentionally violate design guidelines.
In these cases, one might expect to more objectively identify the preferred screen, since the alternative screen would contain a defect.
We found this case well captured by our formulation, as shown by our models' higher performance on the \textsc{JitterWeb} test data.
In contrast, examples from \textsc{BetterApp} are more representative of \textit{design preferences}. Many of its samples were real-world apps, which are often designed professionally and less likely to contain design defects.
To verify this, we analyzed design quality performance on the subset of synthetic, LLM-generated screens within \textsc{BetterApp} and compared it with the app screens from VINS.
Many of the LLM-generated screens (as shown in Figure \ref{fig:example_search}) contain design defects, which potentially led to easier comparisons.
UIClip's accuracy for rating the quality of synthetic screens (67.65\%) was much higher than for real app screens (57.89\%).
This trend was true for almost all other tested models, where the average of all models' accuracy on synthetic screens (56.05\%) was higher than real apps (51.50\%).
It is also possible that UIClip is not trained to detect certain types of design defects present in real-world apps, \textit{e.g.,} violations that require the semantic understanding of content, such as information flow hierarchy.

Besides the nature of design defects in real-world apps, design preferences could also vary by person.
For example, it is reasonable to expect that someone who frequently uses iOS devices might feel more familiar with iOS screenshots over Android ones.
The design language of the same platform can change over time, causing corresponding shifts in user perception, \textit{e.g.,} some screenshots in VINS were from the older Rico dataset \cite{deka2017rico}.
% design defects are easier to detect
% design preference is hard and subjective, it can change from person to person and even within the same person
While UIClip is currently not designed to support more personalized use cases, we envision that it could be finetuned with user-provided preference paired~\cite{gajos2005preference} or augmented so that it incorporates platform-specific design into its prompt, \textit{e.g.,} ``android material design screenshot. well-designed.''
% influenced by many external factors

\subsection{Supporting UI Design Applications}
% could be possible to evaluate layout from a scaled down image
% limitations about multiple patches, scaling
% cropping vs padding
% small window size
% limitations of inference algorithm. not able to localize
% cannot give detailed design feedback
% limitations of each training objective. how to combine the two objectives?
% talka bout single screen limitation
% integrating crawling and app suage
Related to our problem formulation is the types of design-assistance tasks that UIClip can support.
Despite only producing a numerical score as output, we introduced inference techniques that extend beyond simple UI scoring and allow for a limited generation of natural language design suggestions.
We developed three example applications that demonstrate how UIClip could facilitate some forms of automated design assistance.
While we did not conduct formal user evaluations of our example applications, we built these applications based on validated systems described in the literature, which suggest they would provide value to users.

% we support design suggestions, but cannot localize
% talk about free form language from models.
However, there are many types of design assistance that are not yet supported by UIClip.
For example, while UIClip can infer the presence of design defects in a screenshot, there is currently no straightforward method to localize them (\textit{e.g.,} footer bar has poor color contrast).
We believe this capability is important for practical use since it provides cues for designers to address the detected flaws. One promising approach, previously applied to other visual design tasks~\cite{schoop2022predicting}, is to augment our current model with model explainability frameworks to understand which parts of the image contribute to predictions.
Future iterations of the UIClip could also be trained on sub-windows of a UI for finer-grain inference of fault location, similar to how object detection architectures work.
Finally, UIClip could be fine-tuned with more detailed natural language descriptions that associate spatial information with predicted flaws (\textit{e.g.,} ``bad color contrast in footer bar'') or even provide suggested fixes (\textit{e.g.,} ``bad color contrast. make footer darker''), although this would necessitate a more complex inference algorithm.

Recent trends in machine learning suggest that model architectures that generate free-form text can be more easily scaled and provide more flexible feedback. 
Our evaluation found that VLMs generally performed poorly, and prior work suggests that LLMs are prone to providing irrelevant design suggestions~\cite{duan2024generating}.
A qualitative assessment of current LLM responses (Figure \ref{fig:gpt-example}) suggests that current LLMs produce realistic-sounding but inaccurate reasoning.
However, we believe that our work could be useful in improving foundation models such as LVLMs.
For example, the UIClip model could be used as a ``reward model'' that guides their UI generation (Section \ref{sec:uigeneration}) and design assessment capabilities. 
Furthermore, our datasets could be reformatted and used to fine-tune LVLMs for UI-related tasks~\cite{li2022spotlight}.

Finally, most machine learning models, including UIClip and LVLMs are limited in that they can only process a single state of the UI (\textit{i.e.,} a screenshot) when responding to text prompts.
Because of this limitation, our current approach focuses on assessing the visual design of a single screen using the CRAP visual design principles.
A more holistic evaluation of both UI design and usability depends on a deeper understanding of interface functionality and app navigation flows, which requires both observation and interaction.
To support this, we envision that models like UIClip, could be integrated into interactive systems, such as crawlers, that can interact with and explore different parts of an entire app \cite{wu2023never,swearngin2023towards}.

% In summary, UIClip represents a preliminary step in we see many avenues for applying our model and datasets to computationally support UI design, which is a challenging yet important task.
% \textit{Evaluation.}
% we didn't evaluate example apps with real designers, but are primarily based on existing examples found in the literature

% talk about different losses and losses used in baselines

% i think a natural question a reader will have when reading our paper is: will vllms be able to get better at this naturally or is there some need for this approach
% what is generalizable from our approach for other models?

% worried about speculating too much about a lot of evidence to back up at least the hypothesis that llms doesn't have the right knowledge. matter of mapping them. visual models.
% contribution of my paper is to take a known visual model that we show does not have much ability to understand in a UI. and get it to incorporate into its knowledge more knowledge about UIs. so it can reason in a more informed way.

% check if the CLIP model got worse on standard benchmarks.
% probably should have ran humaneval

% \subsection{Limitations \& Future Work}
% cannot explore whole app

% the process probably overweights i like it vs more a principle evaluation.
% they see something, they pick something ,th ey justify it
% maybe we should have evaluated the interface before making a choice
% we used this approach because it's been used in previous papers, stolen directly from ml literature
% gajos paper on prefernce 

% \textit{Modeling Approach.}

\section{Conclusion}
In this paper, we introduce a computational model called UIClip that automatically assesses various aspects of UI design: \textit{i)} design quality, \textit{ii)} UI relevance, and \textit{iii)} is capable of generating design suggestions based on predicted defects. Our model is trained from a large-scale dataset of 2.3 million UIs that we collected and augmented with synthetic and human ratings of design quality. In an evaluation with several strong baselines, we demonstrate our model's performance in UI design understanding in our three target UI tasks, showing that UIClip outperforms all other baselines in all tasks. Finally, we introduce three example applications that demonstrate how UIClip can facilitate novel applications through its automated design assessment capabilities: \textit{i)} UI code generation, \textit{ii)} UI design tips generation, and \textit{iii)} quality-aware UI example search. Overall, our work demonstrates the process of effectively encoding design awareness into the computational models.
% \begin{acks}
% This is the acknowledgements section.
% \end{acks}

\bibliographystyle{ACM-Reference-Format}
\bibliography{sample-base}

\appendix
\section{Hyperparameters}
Table \ref{tab:hyperparameters} provides hyperparameters for various models and algorithms used in our paper. Our CLIP training hyperparameters were based on values from the original CLIP paper~\cite{radford2021learning}, and were manually adjusted to fit on our hardware and based on performance observations.
\begin{table}
\caption{Hyperparameters of models and algorithms used in our paper.}
\small
\begin{tabular}{@{}lll@{}}
\toprule
\textbf{Algorithm} & \textbf{Hyperparam.}      & \textbf{Value}                     \\ \midrule
UIClip (JitterWeb pretraining)         & Batch Size          & 128                         \\
                & Epochs              & 1                         \\
                & Learning Rate       & $5e-7$                      \\
                & Weight Decay               & 0.2                        \\
                & Adam Beta 1               & 0.9                      \\
                & Adam Beta 2         & 0.98                       \\
                & Adam Epsilon           & $1e-6$                        \\                     \\
UIClip (JitterWeb pairwise, BetterApp)         & Batch Size          & 256                         \\
                & Epochs              & 1                         \\
                & Learning Rate       & $5e-7$                      \\
                & Weight Decay               & 0.2                        \\
                & Adam Beta 1               & 0.9                      \\
                & Adam Beta 2         & 0.98                       \\
                & Adam Epsilon           & $1e-6$                        \\                     \\                

BetterApp DBSCAN   & Epsilon         & 0.1                      \\ 
  & Min samples         & 5                      \\ 
  & Metric         & Cosine Similarity                      \\ \bottomrule
\end{tabular}
\label{tab:hyperparameters}
\end{table}

\section{Large Vision-Language Model Prompt}
Below, we provide the prompt that was used to evaluate UI screenshots in our quantitative study. The prompt below is used to evaluate a screenshot of an e-commerce application and contains the same description of CRAP guidelines that we gave to human raters.
\begin{lstlisting}
This image contains two screenshots of user interfaces stacked horizontally (left and right). Both UIs can be described by the description:

An e-commerce application interface

Recall the CRAP guidelines for visual design, which stands for contrast, repetition, alignment, and proximity.

The C.R.A.P principles, coined by Robin Patricia Williams in her non-designers' design book, are a set of guidelines aimed at improving the visual appeal and effectiveness of graphic designs. These principles are essential for creating visually appealing and user-friendly designs. CRAP stands for:
Contrast: This principle suggests that elements that are not the same should be very different so that they stand out. Using contrast can attract the viewer's attention and help organize information. It can be applied through variations in color, size, typeface, and other visual elements.
Repetition: Repetition involves repeating some aspect of the design throughout the entire piece. This can include the consistent use of colors, fonts, and logos, which helps to create a cohesive and harmonious look. Repetition strengthens a design by tying together individual elements and can enhance the overall sense of unity.
Alignment: Every element should have a visual connection with something else on the page. This doesn't mean that elements always need to be in a straight line, but rather that they should be visually connected in a way that makes the entire design appear well organized. Proper alignment eliminates disorder, connects elements, and creates a visually logical structure.
Proximity: Items that relate to each other should be grouped together, which helps in organizing information and reducing clutter. By effectively grouping related elements, the design becomes easier to comprehend, and relationships between elements become clearer to the viewer. Proximity can also help in creating focal points in a design.

Based on these guidelines, provide a response that indicates which UI screenshot is better designed. The first part of your response should contain one of two choices: 'left', 'right.' The second part of your response should contain a comma-separated list of which CRAP principles (if any) are most relevant to your choice. Do not provide explanations, and separate the first and second part of your response with a new line.
\end{lstlisting}
% \section{ChatGPT Output Example}
% \begin{figure*}[!t]
%   \centering
%   \includegraphics[width=1.0\textwidth]{samples/figures/chatgpt_example.png}
%   \caption{foo}\label{fig:chatgpt_example}
%   \Description{}
% \end{figure*}
% \section{Model Hyperparameters}
% this is foo

% \section{Evaluation Guidelines}

% \section{Prompts}
% \subsection{UIClip Training}
% Pix2Struct captioning

% paraphrasing prompt

% UIClip training prompt
% \subsection{UI Code Generator Web Application}
% Generation prompt

% Image asset generation prompt
\end{document}